\documentclass[preprint,showpacs,preprintnumbers,amsmath,amssymb]{revtex4}

\usepackage{graphicx}% Include figure files
\usepackage{dcolumn}% Align table columns on decimal point
\usepackage{bm}% bold math

\begin{document}

\preprint{APS/123-QED}

\title{2PN Light Propagation in the Scalar-Tensor Theory: an $N$-Point-Masses Case}% Force line breaks with \\

\author{Xue-Mei Deng$^{1,3}$}
 \email{xmd@pmo.ac.cn}%Lines break automatically or can be forced with \\
 \author{Yi Xie$^{2,3}$}
\email{yixie@nju.edu.cn}
 \affiliation{%
$^{1}$Purple Mountain Observatory, Chinese Academy of Sciences, Nanjing 210008, China
\\
$^{2}$Department of Astronomy, Nanjing University, Nanjing 210093, China
\\
$^{3}$Key Laboratory of Modern Astronomy and Astrophysics, Nanjing University, Ministry of Education, Nanjing 210093, China
}%

\date{\today}% It is always \today, today,

\begin{abstract}
	Within the framework of the scalar-tensor theory (STT), its second post-Newtonian (2PN) approximation is obtained with Chandrasekhar's approach. By focusing on an $N$-point-masses system as the first step, we reduce the metric to its 2PN form for light propagation. Unlike previous works, at 2PN order, we abandon the hierarchized hypothesis and do not assume two parametrized post-Newtonian (PPN) parameters $\gamma$ and $\beta$ to be unity. We find that although there exist $\gamma$ and $\beta$ in the 2PN metric, only $\gamma$ appears in the 2PN equations of light. As a simple example for applications, a gauge-invariant angle between the directions of two incoming photons for a differential measurement is investigated after the light trajectory is solved in a static and spherically symmetric spacetime. It shows the deviation from the general relativity (GR) $\delta\theta_{\mathrm{STT}}$ does \emph{not} depend on $\beta$ even at 2PN level in this circumstance, which is consistent  with previous results. A more complicated application is light deflection in a 2-point-masses system. We consider a case that the light propagation time is much less than the time scale of its orbital motion and thus treat it as a static system. The 2-body effect at 2PN level originating from relaxing the hierarchized hypothesis is calculated. Our analysis shows the 2PN 2-body effect in the Solar System is one order of magnitude less than future $\sim 1$ nas experiments, while this effect could be comparable with 1PN component of $\delta\theta_{\mathrm{STT}}$ in a binary system with two Sun-like stars and separation by $\sim 0.1$ AU if an experiment would be able to measure $\gamma-1$ down to $\sim 10^{-6}$.
\end{abstract}
\maketitle

\allowdisplaybreaks

\section{Introduction}

Some future space missions, such as the Laser Astrometric Test Of Relativity (LATOR) \cite{tur04,tur04b}, the Phobos Laser Ranging (PLR) \cite{plr}, the Beyond Einstein Advanced Coherent Optical Network (BEACON) \cite{tur09}, the T\'{e}l\'{e}m\'{e}trie InterPlan\'{e}taire Optique (TIPO) \cite{tipo}, the Astrodynamical Space Test of Relativity using Optical Devices (ASTROD) \cite{astrod} and the Search for Anomalous Gravitation using Atomic Sensors (SAGAS) \cite{sagas}, will measure distances of laser links and angles among these links with unprecedented precision. As a sensitive and useful tool in gravitational physics, especially for some high order effects, the propagation of light carries lots of information about the nature of spacetime and plays an important role in high-precision experiments and measurements.

Thus, from the practical and theoretical aspects, it motivates us to investigate a 2PN light propagation model at the $c^{-4}$ level in the framework of the scalar-tensor theory for a gravitational $N$-point-masses system.

\subsection{The necessity of 2PN $c^{-4}$ terms}

For the precision of distance measurements, TIPO could achieve millimeter level \cite{tipo} and BEACON could be even higher, reaching $0.1$ nanometer(nm) \cite{tur09}. As explained in Ref. \cite{min09} (in Sec. V),  a complete metric of the Solar System up to $c^{-4}$ level is demanded for modeling the light propagation in those experiments.  For the angle measurements, the precision of LATOR could achieve nano-arcsecond (nas) level or higher \cite{tur04,tur04b}. It would require $c^{-4}$ terms in the metric to calculate the deflection of light ray, because when light grazes solar limb the 2PN deflection contributed by the Sun is $G^{2}m^{2}_{\odot}/(c^{4}R^{2}_{\odot})\approx10^{-12} \sim $ micro-arcsecond ($\mu$as), which is much larger than the threshold of LATOR. The corrections of alternative theories of gravity to this effect at 2PN order would be at least several orders of magnitude less than $\sim \mu$as. However, these corrections perhaps still need to be considered for future experiments.

Several authors have obtained the 2PN metric for the general relativity (GR) and alternative theories of gravity. Chandrasekhar and Nutku first calculated the 2PN metric and equations of hydrodynamics in GR \cite{b50}. A field theory approach is also employed to derive the 2PN Lagrangian for the equations of motion of $N$ bodies in the multi-scalar-tensor theory without solving the metric \cite{b4}. With introducing an intermediate-range gravity term, the 2PN approximation of the scalar-tensor theory was obtained \cite{xie09} in which the energy-momentum tensor was expressed by using the invariant density \cite{fock}. The 2PN approximation of Einstein-aether theory \cite{b33} was deduced in the form of both superpotentials and an $N$-point-masses. More recently, the IAU2000 resolutions are extended to include all the $c^{-4}$ terms for the requirements from some space missions \cite{min09}. Most of such works express the 2PN metric in terms of superpotentials without definition of the masses and multipole moments of the bodies which is not trivial in the post-Newtonian order especially for the self-gravitating bodies in an $N$-massive-bodies system.

An explicit application of them is to model light propagation in the near zone of a gravitational system into the level of the next leading order. The 2PN light deflection for one body in the standard parametrized post-Newtonian (PPN) formalism \cite{b27,b28} was studied \cite{eps80,ric82} by introducing another parameter ($\epsilon$ or $\Lambda$) at $c^{-4}$ of $g_{ij}$, the space-space component of metric, under the isotropic gauge. The 2PN light propagation in the Schwarzschild spacetime was detailedly researched in three gauges (standard, harmonic and isotropic) through introducing two coordinate parameters \cite{bru91}. Ref. \cite{nov96} discussed the significance of some experiments to observe 2PN light effects for one body in the isotropic gauge. A practical relativistic model for the 2PN light propagation was developed in the harmonic gauge of general relativity \cite{kli}, in which the 1PN contributions from the rectilinear and uniform $N$-body and the 2PN contributions \emph{only} from the Sun were considered. The light propagation in 2PN framework of a stationary gravitational field for the Schwarzschild metric (one body) was formulated in harmonic gauge with introducing one parameter $\epsilon$ at spatial isotropic and anisotropic terms of $c^{-4}$ for $g_{ij}$ \cite{kz}. The 2PN deflection of light in a spherically symmetric body was discussed in GR in three gauges (standard, harmonic and isotropic) \cite{bod03}. All of above works mentioned are only considering one body in the 2PN order. The 2PN gravitational redshift in GR was derived \cite{xu} by superpotentials.  Under the $f(R)$ theory, the 2PN weak lensing was explored in the isotropic gauge \cite{cal11}.

\subsection{The reason for a scalar-tensor theory}

Although Einstein's general relativity has passed nearly all the tests in the Solar System, alternative theories of gravity are still required for deeper understanding of the nature of spacetime and for testing possible violations of the Einstein equivalence principle (EEP) in the forthcoming more precision level \cite{tur08}. In order to testing and distinguishing alternative gravitational theories, the PPN formalism introduces ten parameters in a post-Newtonian metric to include various gravity theories \cite{b27,b28}. However, the PPN formalism is only restricted to the 1PN approximation. Some authors (e.g. \cite{b18,b25,b16}) discussed how to parameterize the 2PN metric. To extend the PPN formalism, one possible way might be to derive 2PN metric for various gravity theories and then find the independent superpotentials in the 2PN level. After that, parameters in the 2PN order could be introduced and endowed with some meanings. However, in this paper, we only focus on one gravitational theory and discuss its parameters.

Among these alternative theories, the most eminent case is a scalar-tensor theory (STT) because it is the simplest and most natural way to modify GR. Many modern theories, such as the extra-dimensional theory, the string theory, the braneworld and the noncommutative geometry, which try to unify gravity and microscopic physics or to explain the dark energy in cosmology, demand a scalar field besides the metric tensor (see \cite{fm} for a review). Especially, Damour and Nordtvedt proposed \cite{dam93} that the deviation of the PPN parameter $\gamma$ from $1$, presenting the discrepancy between STT and GR, might range from $\sim 10^{-7}$ to $\sim 10^{-5}$. The contribution of this deviation in light deflection at 2PN order could be as large as $(\gamma-1)G^{2}m^{2}_{\odot}/(c^{4}R^{2}_{\odot})\approx  10^{-16}\sim 0.02$ nas, which might be important for future detection as well. Therefore, we shall work in this paper with the scalar-tensor theory of gravity.

Motivated by forthcoming space experiments involving propagation of light in the Solar System, some researchers had studied 2PN light propagation in STT.  In the case of a one-body system, 2PN light deflection and light propagation were reported in Refs. \cite{p1} and \cite{p2}. As a more comprehensive work, the scalar-tensor propagation of light was investigated in Ref. \cite{min10}, in which it neglected the differences of PPN parameters $\gamma$ and $\beta$ from $1$ at $c^{-4}$ level according to present experiment results and used the hierarchized hypothesis so that only Sun's contribution remained at $c^{-4}$ level while influences from Sun and planets were all included at $c^{-2}$ and $c^{-3}$ terms.

To make our results fully based on STT, we will keep $\gamma$ and $\beta$ at $c^{-4}$ level which requires a PPN definition of mass, such as Eq.(\ref{tildemu}).

\subsection{The extension to an $N$-point-masses system}

As metioned in Ref. \cite{min10},  the hierarchized hypothesis has its own limitation. In the practical point of view, by abandoning this hypothesis which means extending to an $N$-body system, the solution of light equations might improve the accuracy which could be achieved \cite{min10}. As a rude estimation, one coupling term in 2PN metric $\sum_{a}\sum_{b\neq a}G^{2}m_{a}m_{b}/(c^{4}r_{a}r_{b})$, where $r_{a}=|\bm{x}-\bm{y}_{a}|$, $r_{b}=|\bm{x}-\bm{y}_{b}|$ and the trajectories of the $a$-th and $b$-th masses are respectively represented by $\bm{y}_{a}(t)$ and $\bm{y}_{b}(t)$, could reach  $\sim 10^{-16} $ (equivalent to $\sim 0.1$ nas), if we consider the Solar System (a hierarchic system) that $a$-th mass is the Sun and the $b$-th mass is Jupiter when the light grazes Jupiter's limb. If the influence to the background light sources by a binary system with comparable masses are considered, the coupling term in 2PN might raise larger contributions. Extension to an $N$-body system would also be helpful to determine which terms have to be included for a specific mission and could be a good test-bed to evaluate the accuracy of the hierarchized model \cite{min10}.

In the theoretical point of view, it is a natural development to build a 2PN theory of a gravitational $N$-body system for its dynamics and light propagation within it which might show some subtle but interesting effects due to the non-linearity in the 2PN order. In principle, the 2PN light propagation under the STT should include relativistic multipolar moments of each body in the $N$-body system. 1PN global and local metrics with the definitions of the multipolar moments and spins in STT have been given in Refs. \cite{b19,xie10}.

Therefore, in this paper, we consider 2PN light propagation in an $N$-point-masses system under STT as our first step.

In what follows, our conventions and notations generally
follow those of Ref. \cite{33}. The metric signature is $(-,+,+,+)$.
Greek indices take the values from 0 to 3, while Latin
indices take the values from 1 to 3. A comma denotes a
partial derivative, and dot over a quantity denotes a
derivative with respect to time. Bold letters denote spatial vectors. The plot of
this paper is as follows. In the Sec. II, the 2PN metric and equation of light in the scalar-tensor theory
are given. Subsequently, in Sec. III, we reduce the results of Sec. II to a 2PN metric of an $N$-point-masses system for light propagation by following the method used in Ref. \cite{b20}.
And the 2PN light equation is obtained in this condition.
In Sec. IV, applications will be given. In Sec. \ref{case1pms}, we derive the 2PN light-ray trajectory and deflection in a static, spherically symmetric spacetime
and study the parameters in 2PN terms by comparison among the scalar-tensor theory and others. In Sec. \ref{case2pms}, we calculate the light deflection angle caused by a 2-point-masses system and estimate the magnitudes of effects for different cases.
Finally, the conclusion and discussion are outlined in Sec. V.

\section[]{2PN metric in the scalar-tensor theory and equations of light}

\subsection[]{2PN metric in the scalar-tensor theory}
The action for the scalar-tensor theory based on \cite{b19} reads
\begin{equation}
    \label{lgst2}
    S=\frac{c^3}{16\pi}\int \bigg(\phi R -\frac{\theta(\phi)}{\phi}\phi^{,\sigma}\phi_{,\sigma}
    -\frac{16\pi}{c^4}\mathcal{L}_I(g_{\mu \nu},\Psi) \bigg)\sqrt{-g}\,\mathrm{d}^4x,
\end{equation}
where $c$ is the speed of light, $\theta(\phi)$ is an arbitrary coupling
function of the scalar field $\phi$. $R$ and
$g=\mathrm{det}(g_{\mu\nu})$ denote respectively the Ricci scalar
and the determinant of the metric tensor $g_{\mu\nu}$. The matter field is
denoted by $\Psi$. From Eq. (\ref{lgst2}), we can see
the matter fields $\Psi$ only interact with the metric field (namely, $g_{\mu\nu}$). This
means the trajectory of a free-fall test particle only depends on the
spacetime geometry so that it satisfies EEP. Although violations of EEP at galactic and
cosmological scales can not be ruled out, we still focus on the
scalar-tensor theory satisfied EEP in this paper.

Variation of the action (\ref{lgst2}) with respect to $g^{\alpha
\beta}$ has
\begin{equation}
    \label{feg}
    R_{\mu \nu}  = \frac{8\pi}{\phi c^2}\bigg(T_{\mu \nu}-\frac{1}{2}g_{\mu \nu}T\bigg)
    +\frac{\theta(\phi)}{\phi^2}\phi_{,\mu}\phi_{,\nu}+\frac{1}{\phi}\bigg(\phi_{;\mu\nu}+\frac{1}{2}g_{\mu
    \nu}\Box_g\phi\bigg),
\end{equation}
where $\Box_g(\cdot)=(\cdot)_{;\mu \nu}g^{\mu \nu}$,
$T_{\mu \nu}$ is the stress-energy-momentum tensor of matter which is defined by
\begin{equation}
    \label{}
    \frac{c^2}{2}\sqrt{-g}T_{\mu\nu}\equiv\frac{\partial(\sqrt{-g}\mathcal{L}_I)}{\partial g^{\mu\nu}}-\frac{\partial}{\partial x^\alpha}\frac{\partial(\sqrt{-g}\mathcal{L}_I)}{\partial g^{\mu\nu}_{,\alpha}},
\end{equation}
and $T$ is the trace of $T^{\mu\nu}$. Following Refs. \cite{b46,b47},
the mass, current, and stress densities can be defined as
\begin{eqnarray}
\label{sigma}
\sigma&\equiv& T^{00}+T^{ii},\\
\label{sigmai}
\sigma_{i}&\equiv& cT^{0i},\\
\label{sigmaij}
\sigma_{ij}&\equiv& c^{2}T^{ij}.
\end{eqnarray}
Variation of the action (\ref{lgst2}) with respect to $\phi$ yields
\begin{equation}
    \label{fep}
    \Box_g\phi=\frac{1}{3+2\theta(\phi)}\bigg(\frac{8\pi}{c^2}T-\phi_{,\alpha}\phi^{,\alpha}\frac{\mathrm{d}\theta}{\mathrm{d}\phi}\bigg).
\end{equation}

Based on Ref. \cite{b19}, the scalar field is expanded as its background
value $\phi_{0}$ as follows
\begin{equation}
    \label{scalarpert}
    \phi=\phi_0(1+\zeta),
\end{equation}
where $\zeta$ is a dimensionless perturbation of the background
value $\phi_{0}$. Especially, decomposition of the coupling function
$\theta(\phi)$ reads
\begin{eqnarray}
\label{scalarpert1}
\theta(\phi)=\omega_{0}+\omega_{1}\zeta+\frac{1}{2}\omega_{2}\zeta^{2}+\ldots,
\end{eqnarray}
where $\omega_{0}\equiv\theta(\phi_{0})$, $\omega_{1}\equiv(\mathrm{d}\theta/\mathrm{d}\zeta)_{\phi=\phi_{0}}$, $\ldots$,
$\omega_{n}\equiv(\mathrm{d}^{n}\theta/\mathrm{d}\zeta^{n})_{\phi=\phi_{0}}$, which can lead to the gauge-invariant definitions of PPN parameters $\gamma$ and $\beta$  (see Eqs. (\ref{gamma}) and (\ref{beta}) in Appendix A)
\begin{eqnarray}
    \label{gamma}
    \gamma & \equiv & \frac{\omega_0+1}{\omega_0+2},\nonumber\\
    \label{beta}
    \beta & \equiv & 1+\frac{\omega_1}{(2\omega_0+3)(2\omega_0+4)^2}.\nonumber
\end{eqnarray}

By using Chandrasekhar's approach \cite{b14,b50}, we deal with the
theory in the form of a Taylor expansion in the parameter
$\varepsilon\equiv1/c$. The expansions
of the metric $g_{\mu\nu}$ and the scalar field to the second order
have the forms of
\begin{eqnarray}
    \label{PPNmetric1}
    g_{00} & = & -1+\varepsilon^2 N+\varepsilon^4 L+\varepsilon^6 Q+\mathcal{O}(7),\\
    \label{PPNmetric2}
    g_{0i} & = & \varepsilon^3L_{i}+\varepsilon^5 Q_{i}+\mathcal{O}(6),\\
    \label{PPNmetric3}
    g_{ij} & = & \delta_{ij}+\varepsilon^2H_{ij}+\varepsilon^4Q_{ij}+\mathcal{O}(5),\\
    \label{ezeta}\zeta & = & \varepsilon^2 \overset{(2)}{\zeta}+\varepsilon^4\overset{(4)}{\zeta}+\varepsilon^6\overset{(6)}{\zeta}+\mathcal{O}(7),
\end{eqnarray}
where $O(n)$ means of the order $\varepsilon^{n}$.

According to Eqs. (\ref{scalarpert}), (\ref{scalarpert1}) and
(\ref{ezeta}), we derive
\begin{eqnarray}
    \label{theta(phi)}
    \theta(\phi) & = & \omega_0+\varepsilon^2\omega_1\overset{(2)}{\zeta}+\varepsilon^4\bigg(\frac{1}{2}\omega_2\overset{(2)}{\zeta}^2+\omega_1\overset{(4)}{\zeta}\bigg)
    +\mathcal{O}(5),\\
    \label{dthetadphi}
    \frac{\mathrm{d}\theta}{\mathrm{d}\phi} & = &
    \frac{1}{\phi_0}\bigg(\omega_1+\varepsilon^2\omega_2\overset{(2)}{\zeta}+\mathcal{O}(4)\bigg).
\end{eqnarray}

We use the gauge condition imposed on the component of the metric
tensor proposed by Kopeikin $\&$ Vlasov \cite{b19} as follows:
\begin{equation}
    \label{KVgauge}
    \bigg(\frac{\phi}{\phi_0}\sqrt{-g}g^{\mu \nu}\bigg)_{,\nu}=0.
\end{equation}
Based on fields equations of Eqs. (\ref{feg}) and (\ref{fep}), we
obtain the evolution equations of the metric coefficients at 2PN order by using the gauge
conditions Eq. (\ref{KVgauge}) (see Appendix A for detail).

\subsection{2PN equations of light}

Generally, for a photon propagating in a spacetime in which Einstein Equivalence Principle (EEP) is valid, the basic equations of light
based on Ref. \cite{b26} are
\begin{eqnarray}
\label{LP1}
0&=&g_{\mu\nu}\frac{\mathrm{d}x^{\mu}}{\mathrm{d}t}\frac{\mathrm{d}x^{\nu}}{\mathrm{d}t},\\
\label{LP2}
\frac{\mathrm{d}^{2}x^{i}}{\mathrm{d}t^{2}}&=&\bigg(\varepsilon\Gamma^{0}_{~\nu\sigma}\frac{\mathrm{d}x^{i}}{\mathrm{d}t}
-\Gamma^{i}_{~\nu\sigma}\bigg)\frac{\mathrm{d}x^{\nu}}{\mathrm{d}t}\frac{\mathrm{d}x^{\sigma}}{\mathrm{d}t},
\end{eqnarray}
where we have replaced affine parameter with coordinate time $t$.
Following Ref. \cite{kz}, assuming $\dot{\bm{x}}=cs\bm{\mu}$ and $\bm{\mu}\cdot\bm{\mu}=1$, we find the expression for $s$ from Eq. (\ref{LP1})
\begin{eqnarray}
\label{s}
s&=&1-\frac{1}{2}\varepsilon^{2}(N+H_{ij}\mu^{i}\mu^{j})-\varepsilon^{3}L_{k}\mu^{k}\nonumber\\
&&+\frac{1}{2}\varepsilon^{4}[-Q_{ij}\mu^{i}\mu^{j}-L+\frac{1}{2}N H_{ij}\mu^{i}\mu^{j}-\frac{1}{4}N^{2}+\frac{3}{4}H_{ij}\mu^{i}\mu^{j}H_{kl}\mu^{k}\mu^{l}].
\end{eqnarray}
Then, by substituting the metric Eqs. (\ref{PPNmetric1})-(\ref{PPNmetric3}) into Eq. (\ref{LP2}),
we obtain the equations of light propagation based on Eq. (\ref{LP2}) as follows
\begin{eqnarray}
\label{LE}
\ddot{x}^{i}&=&\frac{1}{2}N_{,i}+\frac{1}{2}H_{jk,i}\frac{\dot{x}^{j}}{c}\frac{\dot{x}^{k}}{c}-H_{ij,k}\frac{\dot{x}^{j}}{c}\frac{\dot{x}^{k}}{c}
-N_{,j}\frac{\dot{x}^{j}}{c}\frac{\dot{x}^{i}}{c}\nonumber\\
&&+\varepsilon\bigg\{-L_{i,j}\frac{\dot{x}^{j}}{c}+L_{j,i}\frac{\dot{x}^{j}}{c}-\frac{1}{2}N_{,t}\frac{\dot{x}^{i}}{c}
+\frac{1}{2}H_{jk,t}\frac{\dot{x}^{j}}{c}\frac{\dot{x}^{k}}{c}\frac{\dot{x}^{i}}{c}
-H_{ij,t}\frac{\dot{x}^{j}}{c}
-L_{k,j}\frac{\dot{x}^{j}}{c}\frac{\dot{x}^{k}}{c}\frac{\dot{x}^{i}}{c}\bigg\}\nonumber\\
&&+\varepsilon^{2}\bigg\{-\frac{1}{2}H_{ik}N_{,k}-N
N_{,j}\frac{\dot{x}^{i}}{c}\frac{\dot{x}^{j}}{c}
+H_{il}H_{lj,k}\frac{\dot{x}^{j}}{c}\frac{\dot{x}^{k}}{c}-\frac{1}{2}H_{il}H_{jk,l}\frac{\dot{x}^{j}}{c}\frac{\dot{x}^{k}}{c}\nonumber\\
&&-L_{i,t}+\frac{1}{2}L_{,i}-L_{,j}\frac{\dot{x}^{i}}{c}\frac{\dot{x}^{j}}{c}-Q_{ij,k}\frac{\dot{x}^{j}}{c}\frac{\dot{x}^{k}}{c}
+\frac{1}{2}Q_{jk,i}\frac{\dot{x}^{j}}{c}\frac{\dot{x}^{k}}{c}\bigg\}+\mathcal{O}(3).
\end{eqnarray}
From Eq. (\ref{LE}), we can see that the 2PN metric for light propagation
could be cut off to
\begin{eqnarray}
    \label{}
    g_{00} & = & -1+\varepsilon^2 N+\varepsilon^4 L+\mathcal{O}(5),\\
    \label{}
    g_{0i} & = & \varepsilon^3L_{i}+\mathcal{O}(5),\\
    \label{}
    g_{ij} & = & \delta_{ij}+\varepsilon^2H_{ij}+\varepsilon^4Q_{ij}+\mathcal{O}(5).
\end{eqnarray}
Furthermore, with using the relationship: $H_{ij}=\delta_{ij}\gamma N$ based on Eqs. (\ref{eqHij}) and (\ref{eqV}) in Appendix A and by substituting $\dot{\bm{x}}\cdot\dot{\bm{x}}=c^{2}s^{2}$ and Eq. (\ref{s}) into Eq. (\ref{LE}), the equation of light is simplified as
\begin{eqnarray*}
\ddot{x}^{i}&=&\frac{1}{2}(1+\gamma)N_{,i}-(1+\gamma)N_{,k}\frac{\dot{x}^{k}}{c}\frac{\dot{x}^{i}}{c}\nonumber\\
&&+\varepsilon\bigg\{-L_{i,j}\frac{\dot{x}^{j}}{c}
+L_{j,i}\frac{\dot{x}^{j}}{c}-\frac{1}{2}(1+\gamma)N_{,t}\frac{\dot{x}^{i}}{c}-\frac{1}{2}L_{j,k}\frac{\dot{x}^{j}}{c}\frac{\dot{x}^{k}}{c}\frac{\dot{x}^{i}}{c}-\frac{1}{2}L_{k,j}\frac{\dot{x}^{j}}{c}
\frac{\dot{x}^{k}}{c}\frac{\dot{x}^{i}}{c}\bigg\}\nonumber\\
&&+\varepsilon^{2}\bigg\{-\gamma(1+\gamma)NN_{,i}-L_{i,t}+\frac{1}{2}L_{,i}+(\gamma^{2}-1)NN_{,k}\frac{\dot{x}^{k}}{c}\frac{\dot{x}^{i}}{c}-\frac{1}{2}Q_{ij,k}\frac{\dot{x}^{j}}{c}\frac{\dot{x}^{k}}{c}
\nonumber\\
&&-\frac{1}{2}Q_{ik,j}\frac{\dot{x}^{j}}{c}\frac{\dot{x}^{k}}{c}+\frac{1}{2}Q_{jk,i}\frac{\dot{x}^{j}}{c}\frac{\dot{x}^{k}}{c}
-L_{,j}\frac{\dot{x}^{j}}{c}\frac{\dot{x}^{i}}{c}\bigg\}+\mathcal{O}(3),
\end{eqnarray*}

\section[]{2PN light-ray propagation in a system of $N$ point masses}

\subsection{Solving the metric for 2PN light propagation}

Considering an $N$-body system of nonspinning point
masses as our first move, we follow the notation adopted by Ref. \cite{b20} and use the matter stress energy tensor as follows:
\begin{equation}
c^{2}T^{\mu\nu}(\bm{x},t)=\sum_{a}\mu_{a}(t)\upsilon_{a}^{\mu}\upsilon_{a}^{\nu}\delta(\bm{x}-\bm{y}_{a}(t)),
\end{equation}
where $\delta$ denotes the three-dimensional Dirac distribution, the
trajectory of the $a$-th mass is represented by $\bm{y}_{a}(t)$,
the coordinate velocity of the $a$-th body is
$\bm{v}_{a}=\mathrm{d}\bm{y}_{a}(t)/\mathrm{d}t$ and
$\upsilon_{a}^{\mu}\equiv(c,\bm{v}_{a})$, and $\mu_{a}$ denotes
an effective time-dependent mass of the $a$-th body defined by
\begin{equation}
\mu_{a}(t)=\Bigg(\frac{m_{a}}{\sqrt{gg_{\rho\sigma}\frac{\upsilon_{a}^{\rho}\upsilon_{a}^{\sigma}}{c^{2}}}}\Bigg)_{a},
\end{equation}
where subscript $a$ denotes evaluation at the $a$-th body and $m_{a}$
is the constant Schwarzschild mass. Another useful notation is
\begin{equation}
\tilde{\mu}_{a}(t)=\mu_{a}(t)\bigg[1+\frac{\upsilon_{a}^{2}}{c^{2}}\bigg],
\end{equation}
where $\upsilon_{a}^{2}=\bm{v}_{a}\cdot\bm{v}_{a}$. Both $\mu_{a}$ and
$\tilde{\mu}_{a}$ reduce to the Schwarzschild mass at Newtonian
order: $\mu_{a}=m_{a}+\mathcal {O}(2)$ and
$\tilde{\mu}_{a}=m_{a}+\mathcal {O}(2)$. Then the mass,
current, and stress densities in Eqs. (\ref{sigma}), (\ref{sigmai})
and (\ref{sigmaij}) for the $N$ point masses read
\begin{eqnarray}
\sigma&=&\sum_{a}\tilde{\mu}_{a}\delta(\bm{x}-\bm{y}_{a}(t)),\\
\sigma_{i}&=&\sum_{a}\mu_{a}\upsilon_{a}^{i}\delta(\bm{x}-\bm{y}_{a}(t)),\\
\sigma_{ij}&=&\sum_{a}\mu_{a}\upsilon^{i}_{a}\upsilon^{j}_{a}\delta(\bm{x}-\bm{y}_{a}(t)).
\end{eqnarray}
The next step is to work out $N$, $\overset{(2)}{\zeta}$ and $H_{ij}$ in 1PN approximation
as
\begin{eqnarray}
\label{NPoint}
N&=&2\Box^{-1}\{-4\pi G\sigma\}\nonumber\\
&=&2\sum_{a}
G\bigg\{\frac{\tilde{\mu}_{a}}{r_{a}}-\varepsilon\partial_{t}(\tilde{\mu}_{a})+\varepsilon^{2}\frac{1}{2}\partial^{2}_{t}(\tilde{\mu}_{a}r_{a})
\bigg\}+\mathcal{O}(3)\nonumber\\
&=&2\sum_{a}\frac{Gm_{a}}{r_{a}}
+\varepsilon^{2}\bigg\{\sum_{a}\frac{Gm_{a}}{r_{a}}\bigg[4v^{2}_{a}-(n_{a}v_{a})^{2}\bigg]
+2(2-3\gamma)\sum_{a}\sum_{b\neq
a}\frac{G^{2}m_{a}m_{b}}{r_{a}r_{ab}}\nonumber\\
&&+\sum_{a}\sum_{b\neq
a}\frac{G^{2}m_{a}m_{b}}{r^{2}_{ab}}(n_{a}n_{ab})\bigg\}+\mathcal{O}(3),
\end{eqnarray}
\begin{eqnarray}
\label{zpoint}
\overset{(2)}{\zeta}&=&(1-\gamma)\sum_{a}\frac{Gm_{a}}{r_{a}}
+\varepsilon^{2}\bigg\{\frac{1}{2}(1-\gamma)\sum_{a}\frac{
Gm_{a}}{r_{a}}\bigg[4v^{2}_{a}-(n_{a}v_{a})^{2}\bigg]
\nonumber\\
&&+(1-\gamma)(2-3\gamma)\sum_{a}\sum_{b\neq a}\frac{
G^{2}m_{a}m_{b}}{r_{a}r_{ab}}+\frac{1}{2}(1-\gamma)\sum_{a}\sum_{b\neq
a}\frac{G^{2}m_{a}m_{b}}{r^{2}_{ab}}(n_{a}n_{ab})\bigg\}\nonumber\\
&&+\mathcal{O}(3),
\end{eqnarray}
\begin{eqnarray}
\label{HPoint}
H_{ij}&=&2\delta_{ij}\gamma\sum_{a}\frac{Gm_{a}}{r_{a}}
+\varepsilon^{2}\delta_{ij}\gamma\bigg\{\sum_{a}\frac{Gm_{a}}{r_{a}}\bigg[4v^{2}_{a}-(n_{a}v_{a})^{2}\bigg]
+2(2-3\gamma)\sum_{a}\sum_{b\neq
a}\frac{G^{2}m_{a}m_{b}}{r_{a}r_{ab}}\nonumber\\
&&+\sum_{a}\sum_{b\neq
a}\frac{G^{2}m_{a}m_{b}}{r^{2}_{ab}}(n_{a}n_{ab})\bigg\}+\mathcal{O}(3),
\end{eqnarray}
by the relation of $\tilde{\mu}_{a}$ that
\begin{eqnarray}
\label{tildemu}
\tilde{\mu}_{a}&=&m_{a}\bigg\{1+\varepsilon^{2}\bigg[\bigg(N-\frac{1}{2}\delta_{ij}H_{ij}\bigg)_{a}
+\frac{3}{2}v^{2}_{a}\bigg]+\mathcal {O}(4)\bigg\}\nonumber\\
&=&m_{a}\bigg\{1+\varepsilon^{2}\bigg[(2-3\gamma)\sum_{b\neq
a}\frac{Gm_{b}}{r_{ab}}+\frac{3}{2}v^{2}_{a}\bigg]+\mathcal {O}(4)\bigg\},
\end{eqnarray}
where $r_{a}=|\bm{x}-\bm{y}_{a}|$ and
$r_{ab}=|\bm{y}_{a}-\bm{y}_{b}|$. Scalar products are denoted with parentheses such as $(n_{a}v_{a})=\bm{n}_{a}\cdot\bm{v}_{a}$,
$\bm{n}_{a}=(\bm{x}-\bm{y}_{a})/r_{a}$ and $\bm{n}_{ab}=(\bm{y}_{a}-\bm{y}_{b})/r_{ab}$.
For $L_{i}$,
\begin{eqnarray}
\label{Lipoint}
L_{i}&=&-2(1+\gamma)\Box^{-1}\{-4\pi G\sigma_{i}\}=\sum_{a}\int\frac{Gm_{a}v^{i}_{a}\delta(\bm{z}-\bm{y}_{a})}{|\bm{x}-\bm{z}|}d^{3}z
=\sum_{a}\frac{Gm_{a}}{r_{a}}v^{i}_{a}+\mathcal{O}(2).
\end{eqnarray}
Based on Eq. (\ref{L}), $L$ can be simplified as
\begin{eqnarray}
\label{Lpoint}
L&=&(3\gamma-2\beta-1)\Box^{-1}\{-4\pi G\sigma\}+2(\gamma-1)\Box^{-1}\{-4\pi G\sigma_{kk}\}-\frac{1}{2}\beta N^{2}\nonumber\\
&=&(3\gamma-2\beta-1)\sum_{a}G\bigg\{\frac{\tilde{\mu}_{a}}{r_{a}}(N)_{a}\bigg\}
+2(\gamma-1)\sum_{a}G\bigg\{\frac{\mu_{a}}{r_{a}}v^{2}_{a}\bigg\}
-\frac{1}{2}\beta N^{2}\nonumber\\
&=&2(\gamma-1)\sum_{a}\frac{Gm_{a}}{r_{a}}v^{2}_{a}-2\beta\sum_{a}\frac{G^{2}m^{2}_{a}}{r^{2}_{a}}+2(3\gamma-2\beta-1)\sum_{a}\sum_{b\neq
a}\frac{G^{2}m_{a}m_{b}}{r_{a}r_{ab}}\nonumber\\
&&-2\beta\sum_{a}\sum_{b\neq a}\frac{G^{2}m_{a}m_{b}}{r_{a}r_{b}}+\mathcal{O}(1),
\end{eqnarray}
with the help of
\begin{eqnarray}
\sum_{a}\frac{G\tilde{\mu}_{a}}{r_{a}}(N)_{a}&=&2\sum_{a}\sum_{b\neq
a}\frac{G^{2}m_{a}m_{b}}{r_{a}r_{ab}}+\mathcal{O}(2),\\
\sum_{a}\frac{G\mu_{a}}{r_{a}}v^{2}_{a}&=&\sum_{a}\frac{Gm_{a}}{r_{a}}v^{2}_{a}+\mathcal{O}(2),\\
-\frac{1}{2}\beta
N^{2}&=&-2\beta\sum_{a}\frac{G^{2}m^{2}_{a}}{r^{2}_{a}}-2\beta\sum_{a}\sum_{b\neq
a}\frac{G^{2}m_{a}m_{b}}{r_{a}r_{b}}+\mathcal{O}(2).
\end{eqnarray}
And
\begin{eqnarray}
\overset{(4)}{\zeta}&=&2(\gamma-1)\sum_{a}\frac{Gm_{a}}{r_{a}}v^{2}_{a}+(1+6\gamma-3\gamma^{2}-4\beta)\sum_{a}\sum_{b\neq
a}\frac{G^{2}m_{a}m_{b}}{r_{a}r_{ab}}\nonumber\\
&&+(\frac{5}{2}-\gamma+\frac{1}{2}\gamma^{2}-2\beta)\sum_{a}\frac{G^{2}m^{2}_{a}}{r^{2}_{a}}
+(\frac{5}{2}-\gamma+\frac{1}{2}\gamma^{2}-2\beta)\sum_{a}\sum_{b\neq
a}\frac{G^{2}m_{a}m_{b}}{r_{a}r_{b}}\nonumber\\
&&+\mathcal{O}(1).
\end{eqnarray}
It follows that we derive the metric $Q_{ij}$ as follows
\begin{eqnarray}
\Box Q_{ij}&=&-8(1+\gamma)\pi
G\sigma_{ij}-\frac{1}{2}(1+\gamma)N_{,i}N_{,j}\nonumber\\
&&+\delta_{ij}\bigg\{8\pi
G\sigma(\frac{3}{2}\gamma-\frac{3}{2}\gamma^{2}-\beta+1)N+16\gamma\pi
G\sigma_{kk}+\frac{1}{2}(\gamma^{2}+\beta-1)\nabla^{2}(N^{2})\bigg\},
\end{eqnarray}
where the quadratic part of potentials in $Q_{ij}$ can be rewritten as
\begin{eqnarray}
N_{,i}N_{,j}&=&4\sum_{a}G^{2}m^{2}_{a}\bigg(\frac{1}{r_{a}}\bigg)_{,i}\bigg(\frac{1}{r_{a}}\bigg)_{,j}+4\sum_{a}\sum_{b\neq
a}G^{2}m_{a}m_{b}\bigg(\frac{1}{r_{a}}\bigg)_{,i}\bigg(\frac{1}{r_{b}}\bigg)_{,j}\nonumber\\
&=&\frac{1}{2}\sum_{a}G^{2}m^{2}_{a}(\partial^{2}_{ij}+\delta_{ij}\Delta)\bigg(\frac{1}{r^{2}_{a}}\bigg)+4\sum_{a}\sum_{b\neq
a}G^{2}m_{a}m_{b}\partial_{ai}\partial_{bj}\bigg(\frac{1}{r_{a}r_{b}}\bigg),
\end{eqnarray}
where $\partial_{ai}$ denotes the partial derivative with respect to $\bm{y}_{a}$. The integral of the self-terms can be readily deduced from
$\Delta\ln r_{a}=1/r^{2}_{a}$ \cite{b20}. On the other hand, the interaction terms are obtained by
\begin{eqnarray}
\Delta \ln S_{ab}=\frac{1}{r_{a}r_{b}},
\end{eqnarray}
where $S_{ab}=r_{a}+r_{b}+r_{ab}$ \cite{fock}. Consequently we solve
\begin{eqnarray}
&&-\frac{1}{2}(1+\gamma)\Box^{-1}[N_{,i}N_{,j}]\nonumber\\
&=&-\frac{1}{4}(1+\gamma)\sum_{a}G^{2}m^{2}_{a}\bigg(\partial^{2}_{ij}\ln
r_{a}+\frac{\delta_{ij}}{r^{2}_{a}}\bigg)-2(1+\gamma)\sum_{a}\sum_{b\neq
a}G^{2}m_{a}m_{b}\partial_{ai}\partial_{bj}\ln S_{ab}\nonumber\\
&=&\frac{1}{2}(1+\gamma)\sum_{a}G^{2}m^{2}_{a}\bigg(\frac{n^{i}_{a}n^{j}_{b}}{r^{2}_{a}}-\frac{\delta_{ij}}{r^{2}_{a}}\bigg)-2(1+\gamma)\sum_{a}
\sum_{b\neq
a}G^{2}m_{a}m_{b}\bigg[\frac{n^{i}_{ab}n^{j}_{ab}-\delta_{ij}}{r_{ab}S_{ab}}\nonumber\\
&&+\frac{(n^{i}_{ab}-n^{i}_{a})(n^{j}_{ab}+n^{j}_{b})}{S^{2}_{ab}}\bigg],
\end{eqnarray}
where two relations that
\begin{eqnarray}
\partial^{2}_{ij}\ln r_{a}=\frac{\delta_{ij}-2n^{i}_{a}n^{j}_{a}}{r^{2}_{a}},
\end{eqnarray}
and
\begin{eqnarray}
\partial_{ai}\partial_{bj}\ln S_{ab}=\frac{n^{i}_{ab}n^{j}_{ab}-\delta_{ij}}{r_{ab}S_{ab}}+\frac{(n^{i}_{ab}-n^{i}_{a})(n^{j}_{ab}+n^{j}_{b})}{S^{2}_{ab}},
\end{eqnarray}
are used.
Then, $Q_{ij}$ can be worked out as
\begin{eqnarray}
\label{Qij}
Q_{ij}&=&2(1+\gamma)\sum_{a}\frac{Gm_{a}}{r_{a}}v^{i}_{a}v^{j}_{a}+\frac{1}{2}(1+\gamma)\sum_{a}\frac{G^{2}m^{2}_{a}}{r^{2}_{a}}n^{i}_{a}n^{j}_{a}-2(1+\gamma)\sum_{a}\sum_{b\neq
a}\frac{G^{2}m_{a}m_{b}}{r_{ab}S_{ab}}n^{i}_{ab}n^{j}_{ab}\nonumber\\
&&-2(1+\gamma)\sum_{a}\sum_{b\neq a}\frac{G^{2}m_{a}m_{b}}{S^{2}_{ab}}(n^{i}_{ab}-n^{i}_{a})(n^{j}_{ab}+n^{j}_{b})\nonumber\\
&&+\delta_{ij}\bigg[-4\gamma\sum_{a}\frac{Gm_{a}}{r_{a}}v^{2}_{a}-2(3\gamma-3\gamma^{2}-2\beta+2)\sum_{a}\sum_{b\neq
a}\frac{G^{2}m_{a}m_{b}}{r_{a}r_{ab}}\nonumber\\
&&+(2\gamma^{2}-\frac{1}{2}\gamma+2\beta-\frac{5}{2})\sum_{a}\frac{G^{2}m^{2}_{a}}{r^{2}_{a}}
+2(1+\gamma)\sum_{a}\sum_{b\neq
a}\frac{G^{2}m_{a}m_{b}}{r_{ab}S_{ab}}\nonumber\\
&&+2(\gamma^{2}+\beta-1)\sum_{a}\sum_{b\neq
a}\frac{G^{2}m_{a}m_{b}}{r_{a}r_{b}}\bigg]+\mathcal{O}(1).
\end{eqnarray}

\subsection{The 2PN metric of light-ray propagation for an $N$-point-masses case}

Collecting all these results together, we have the 2PN metric for light propagation in the scalar-tensor theory as
\begin{eqnarray}
g_{00}&=&-1+\varepsilon^{2}2\sum_{a}\frac{Gm_{a}}{r_{a}}+\varepsilon^{4}\bigg\{-2\beta\sum_{a}\frac{G^{2}m^{2}_{a}}{r^{2}_{a}}
+2(1+\gamma)\sum_{a}\frac{Gm_{a}}{r_{a}}v^{2}_{a}
-\sum_{a}\frac{Gm_{a}}{r_{a}}(n_{a}v_{a})^{2}\nonumber\\
&&+2(2-2\beta-1)\sum_{a}\sum_{b\neq
a}\frac{G^{2}m_{a}m_{b}}{r_{a}r_{ab}}+\sum_{a}\sum_{b\neq
a}\frac{G^{2}m_{a}m_{b}}{r^{2}_{ab}}(n_{ab}n_{a})-2\beta\sum_{a}\sum_{b\neq
a}\frac{G^{2}m_{a}m_{b}}{r_{a}r_{b}}\bigg\}\nonumber\\
&&+\mathcal{O}(5),\\
g_{0i}&=&-\varepsilon^{3}2(1+\gamma)\sum_{a}\frac{Gm_{a}}{r_{a}}v^{i}_{a}+\mathcal{O}(5),\\
g_{ij}&=&\delta_{ij}+\varepsilon^{2}2\gamma\sum_{a}\frac{Gm_{a}}{r_{a}}\delta_{ij}+\varepsilon^{4}\delta_{ij}\bigg\{
(2\gamma^{2}-\frac{1}{2}\gamma+2\beta-\frac{5}{2})\sum_{a}\frac{G^{2}m^{2}_{a}}{r^{2}_{a}}
-\gamma\sum_{a}\frac{Gm_{a}}{r_{a}}(n_{a}v_{a})^{2}\nonumber\\
&&+\sum_{a}\sum_{b\neq
a}G^{2}m_{a}m_{b}\bigg[2(\gamma^{2}+\beta-1)\frac{1}{r_{a}r_{b}}+2(2\beta-\gamma-2)\frac{1}{r_{a}r_{ab}}+\gamma\frac{(n_{a}n_{ab})}{r^{2}_{ab}}+2(1+\gamma)\frac{1}{r_{ab}S_{ab}}\bigg]\bigg\}\nonumber\\
&&+\varepsilon^{4}\bigg\{\frac{1}{2}(1+\gamma)\sum_{a}\frac{G^{2}m^{2}_{a}}{r^{2}_{a}}n^{i}_{a}n^{j}_{a}
+2(1+\gamma)\sum_{a}\frac{Gm_{a}}{r_{a}}v^{i}_{a}v^{j}_{a}\nonumber\\
&&-2(1+\gamma)\sum_{a}\sum_{b\neq
a}G^{2}m_{a}m_{b}\bigg[\frac{n^{i}_{ab}n^{j}_{ab}}{r_{ab}S_{ab}}+\frac{(n^{i}_{ab}-n^{i}_{a})(n^{j}_{ab}+n^{j}_{b})}{S^{2}_{ab}}\bigg]\bigg\}+\mathcal{O}(5).
\end{eqnarray}
If the above result returns to GR ($\gamma=\beta=1$) and two-body case, it coincides with the result of Ref. \cite{b20}. It is worth noting that the parameters at the spatial isotropic ($\delta_{ij}\sum_{a}\frac{G^{2}m^{2}_{a}}{r^{2}_{a}}$) and anisotropic ($\sum_{a}\frac{G^{2}m^{2}_{a}}{r^{4}_{a}}r^{i}_{a}r^{j}_{a}$) parts of $\varepsilon^{4}$ for $g_{ij}$ are respectively $(2\gamma^{2}-\frac{1}{2}\gamma+2\beta-\frac{5}{2})$ and $\frac{1}{2}(1+\gamma)$ in the scalar-tensor theory (STT). For one-body case, different theories impose different values on the above parameters (see Table \ref{gammabeta}). When we take Einstein's general relativity (GR), the parameters coming from two parts are totally unity \cite{kli}. Ref. \cite{b33} shows, for a single body, Einstein-aether theory (AE-theory) has two unequal coefficients of the spatial isotropic and anisotropic part in the 2PN $g_{ij}$ but only one parameter $c_{14}$ presents in them. Recently, Ref. \cite{kz} introduces only one parameter $\epsilon$ to parametrize these two parts. From our calculations, it shows the parametrization with single parameter $\epsilon$ in Ref. \cite{kz} is \emph{not} valid for the STT.
Parametrized 2PN issues are more complicated and were researched to a certain extent by the previous works \cite{b18,b25,b16}, but it is beyond the territory of this paper.
\begin{table}
\caption{\label{gammabeta} Parameters at the spatial isotropic and anisotropic parts of $\varepsilon^{4}$ for $g_{ij}$ in different theories.}
\begin{ruledtabular}
\begin{tabular}{cccccccccc}
Parameter  & GR & AE-theory & Parameter in & STT\\
at & \cite{kli} & \cite{b33} & \cite{kz} & Our work\\
\hline
$\delta_{ij}\frac{G^{2}m^{2}}{r^{2}}$ & 1 &$1+\frac{1}{2}c_{14}$ & $\epsilon$ & $2\gamma^{2}-\frac{1}{2}\gamma+2\beta-\frac{5}{2}$\\
$\frac{G^{2}m^{2}}{r^{4}}r^{i}r^{j}$ & 1 &$1-\frac{1}{2}c_{14}$ & $\epsilon$ & $\frac{1}{2}(1+\gamma)$\\
\end{tabular}
\end{ruledtabular}
\end{table}

\subsection{The light-ray propagation equations}

Substituting metric coefficients $N$ for Eq. (\ref{NPoint}), $H_{ij}$ for Eq. (\ref{HPoint}), $L_{i}$ for Eq. (\ref{Lipoint}), $L$ for Eq. (\ref{Lpoint}) and $Q_{ij}$ for Eq. (\ref{Qij}) into Eq. (\ref{LE}), we have
\begin{eqnarray}
\label{le2}
\ddot{x}^{i}=F^{i}_{1PN}+F^{i}_{2PN}+F^{i}_{v},
\end{eqnarray}
where the 1PN monopole component is
\begin{eqnarray}
F^{i}_{1PN}&=&-(1+\gamma)\sum_{a}\frac{Gm_{a}}{r^{2}_{a}}\bigg[n^{i}_{a}-2\frac{(n_{a}\dot{x})\dot{x}^{i}}{c^{2}}\bigg],
\end{eqnarray}
the 2PN monopole component is
\begin{eqnarray}
F^{i}_{2PN}&=&\varepsilon^{2}(1+\gamma)\sum_{a}\frac{G^{2}m^{2}_{a}}{r^{3}_{a}}\bigg\{2(1+\gamma)n^{i}_{a}
+\frac{(n_{a}\dot{x})^{2}}{c^{2}}n^{i}_{a}
-\frac{(n_{a}\dot{x})\dot{x}^{i}}{c^{2}}\bigg\}\nonumber\\
&&+\varepsilon^{2}(1+\gamma)\sum_{a}\sum_{b\neq a}\frac{G^{2}m_{a}m_{b}}{r_{a}}\bigg\{\frac{1}{r_{a}r_{ab}}n^{i}_{a}
-\frac{1}{2r^{2}_{ab}}(n_{a}n_{ab})n^{i}_{a}
+(1-\gamma)\frac{1}{r_{a}r_{b}}n^{i}_{a}\nonumber\\
&&+(1+3\gamma)\frac{1}{r^{2}_{b}}n^{i}_{b}
-\frac{3}{2r^{2}_{ab}}n^{i}_{ab}
-\frac{2}{r_{a}r_{ab}}\frac{(n_{a}\dot{x})\dot{x}^{i}}{c^{2}}
+\frac{(n_{a}n_{ab})}{r^{2}_{ab}}\frac{(n_{a}\dot{x})\dot{x}^{i}}{c^{2}}
-\frac{1}{r^{2}_{ab}}\frac{(n_{ab}\dot{x})\dot{x}^{i}}{c^{2}}\nonumber\\
&&+2(1-\gamma)\frac{1}{r^{2}_{b}}\frac{(n_{b}\dot{x})\dot{x}^{i}}{c^{2}}
-2(1-\gamma)\frac{1}{r_{a}r_{b}}\frac{(n_{a}\dot{x})\dot{x}^{i}}{c^{2}}\bigg\}\nonumber\\
&&+\varepsilon^{2}(1+\gamma)\sum_{a}\sum_{b\neq a}\frac{G^{2}m_{a}m_{b}}{S^{2}_{ab}}\bigg\{-2\frac{1}{r_{ab}}\frac{(n_{ab}\dot{x})(n_{a}\dot{x})+(n_{ab}\dot{x})(n_{b}\dot{x})}{c^{2}}n^{i}_{ab}\nonumber\\
&&-4\frac{1}{S_{ab}}\frac{[(n_{ab}\dot{x})+(n_{b}\dot{x})][(n_{a}\dot{x})+(n_{b}\dot{x})]}{c^{2}}(n^{i}_{ab}-n^{i}_{a})
+\frac{1}{r_{a}}\frac{(n_{ab}\dot{x})+(n_{b}\dot{x})}{c}\bigg[\frac{(n_{a}\dot{x})}{c}n^{i}_{a}-\frac{\dot{x}^{i}}{c}\bigg]\nonumber\\
&&-2\frac{1}{r_{b}}\bigg[\frac{(n_{b}\dot{x})^{2}}{c^{2}}-1\bigg](n^{i}_{ab}-n^{i}_{a})
+2\frac{1}{r_{ab}}\bigg[\frac{(n_{a}\dot{x})\dot{x}^{i}}{c^{2}}+\frac{(n_{b}\dot{x})\dot{x}^{i}}{c^{2}}\bigg]
+\frac{1}{r_{ab}}\frac{(n_{ab}\dot{x})^{2}}{c^{2}}[n^{i}_{a}+n^{i}_{b}]\nonumber\\
&&+2\frac{1}{S_{ab}}\frac{[(n_{ab}\dot{x})-(n_{a}\dot{x})][(n_{ab}\dot{x})+(n_{b}\dot{x})]}{c^{2}}(n^{i}_{a}+n^{i}_{b})
+\frac{1}{r_{b}}\frac{(n_{ab}\dot{x})-(n_{a}\dot{x})}{c}\bigg[\frac{(n_{b}\dot{x})}{c}n^{i}_{b}-\frac{\dot{x}^{i}}{c}\bigg]\nonumber\\
&&-\frac{1}{r_{ab}}(n^{i}_{a}+n^{i}_{b})\bigg\},
\end{eqnarray}
the influence of  the bodies' orbital motions is
\begin{eqnarray}
F^{i}_{v}&=&-\varepsilon2(1+\gamma)\sum_{a}\frac{Gm_{a}}{r^{2}_{a}}\bigg\{\frac{(n_{a}\dot{x})}{c}v^{i}_{a}-\frac{(v_{a}\dot{x})}{c}n^{i}_{a}
+\bigg[\frac{1}{2}(n_{a}v_{a})+\frac{(v_{a}\dot{x})(n_{a}\dot{x})}{c^{2}}\bigg]\frac{\dot{x}^{i}}{c}\bigg\}\nonumber\\
&&+\varepsilon^{2}(1+\gamma)\sum_{a}\frac{Gm_{a}}{r^{2}_{a}}\bigg\{
-\bigg[v^{2}_{a}-\frac{3}{2}(n_{a}v_{a})^{2}+\frac{(v_{a}\dot{x})^{2}}{c^{2}}\bigg]n^{i}_{a}
+\bigg[(n_{a}v_{a})+2\frac{(v_{a}\dot{x})(n_{a}\dot{x})}{c^{2}}\bigg]v^{i}_{a}\nonumber\\
&&+\bigg[2v^{2}_{a}\frac{(n_{a}\dot{x})}{c}
-3(n_{a}v_{a})^{2}\frac{(n_{a}\dot{x})}{c}
+2(n_{a}v_{a})\frac{(v_{a}\dot{x})}{c}\bigg]\frac{\dot{x}^{i}}{c}
\bigg\}.
\end{eqnarray}
In the case of a one-body system studied by many previous works, we always can construct a reference frame within which the body is static so that the velocity-dependent terms vanish.

\section{Applications in an $N$-point-masses system: $N=1$ and $N=2$}

After obtaining the light-ray propagation equations, we will apply them into a 1-point-mass system ($N=1$) and a 2-point-masses system ($N=2$) to calculate the deflection angles for differential measurements.

\subsection{Deflection angle by a 1-point-mass system}

\label{case1pms}

For a 1-point-mass system, it has a static and spherically symmetric spacetime. The 2PN light deflection angle caused by the central body has been computed in some previous works, such as Ref. \cite{eps80,ric82,bru91,nov96,kli,kz,bod03}. Here, we leave the details of calculation in Appendix \ref{lightrayonebody} and show the final result of the deflection angle $\Delta\theta \equiv \theta-\vartheta_{0}$ as
\begin{eqnarray}
\label{def}
\Delta\theta &=& \bigg(\frac{1+\gamma}{2}\bigg)\frac{4Gm_{a}}{c^{2}d_{a}}\nonumber\\
&&-(1+\gamma)\bigg[2(1+\gamma)-\frac{1}{8}(7+8\gamma)\pi\bigg]\frac{G^{2}m^{2}_{a}}{c^{4}d^{2}_{a}}.
\end{eqnarray}
where $\bm{d}_{a}=\bm{k}\times(\bm{r}_{a}\times\bm{k})$
is an impact parameter to represent the closest distance between the
unperturbed light ray and body $a$, $d_{a}=|\bm{d}_{a}|$. When $\gamma=1$, Eq. (\ref{def}) will reduce to the result of GR \cite{b26}.  And it is identical with the deflection angle reported in Ref.\cite{p2}, which considers 2PN light propagation in a one-body system under the framework of STT.

\begin{table}
\caption{\label{discuss} Summary of 2PN parameters in different theories.}
\begin{ruledtabular}
\begin{tabular}{cccccccccc}
Theories  & Parameters & Parameters in & Parameters in  \\
          &  in metric &  light-ray trajectory & 2PN deflection \\
\hline
GR \cite{bod03}& None & None & $-8+\frac{15}{4}\pi$\\
AE-theory \cite{b33} & $c_{14}$ &$c_{14}$ & $-8+\bigg(\frac{15}{4}+\frac{1}{8}c_{14}\bigg)\pi$\\
From Ref.\cite{kz} &$\gamma,\beta,\epsilon$ & $\gamma,\beta,\epsilon$ & $-2(1+\gamma)^{2}+[2(1+\gamma)-\beta+\frac{3}{4}\epsilon]\pi$\\
STT [our work] &$\gamma,\beta$ & $\gamma$ & $-2(1+\gamma)^{2}+\frac{1}{8}(1+\gamma)(7+8\gamma)\pi$\\
\end{tabular}
\end{ruledtabular}
\end{table}

Table \ref{discuss} lists the summary of 2PN parameters appearing in the metric, light-ray trajectory and deflection angle in (1) GR; (2) AE-theory; (3) Ref.\cite{kz}; (4) STT (our work).
In GR, there is no any parameter. For AE-theory, there is only one parameter $c_{14}$.
From Ref.\cite{kz}, three parameters, $\gamma$, $\beta$ and $\epsilon$, show up.
It is very interesting that only one parameter $\gamma$ appears in 2PN light-ray trajectory and light deflection for the scalar-tensor theory although
there are two parameters $\gamma$ and $\beta$ in 2PN metric for the theory. If we
let $\bar{\gamma}\equiv\gamma-1$, Eq. (\ref{def}) returns to
\begin{equation}
  \label{}
  \Delta\theta = \delta\theta_{\mathrm{GR}}+\delta\theta_{\mathrm{STT}},
\end{equation}
where
\begin{eqnarray}
  \label{thetaGR1be}
  \delta\theta_{\mathrm{GR}} & = & \frac{4Gm_{a}}{c^{2}d_{a}}-(8-\frac{15}{4}\pi)\frac{G^{2}m^{2}_{a}}{c^{4}d_{a}^2}\nonumber\\
  & = & 1.75 \bigg(\frac{m_a}{M_{\odot}}\bigg)\bigg(\frac{R_{\odot}}{d_a}\bigg)\quad \mathrm{arcsecond}\nonumber\\
  & &+3.50\bigg(\frac{m_a}{M_{\odot}}\bigg)^2\bigg(\frac{R_{\odot}}{d_a}\bigg)^2\quad \mu\mathrm{as},
\end{eqnarray}
\begin{eqnarray}
  \label{deltatheta1be}
  \delta\theta_{\mathrm{STT}} & = & \bar{\gamma}\frac{2Gm_{a}}{c^{2}d_{a}}
-\bar{\gamma}(8-\frac{31}{8}\pi)\frac{G^{2}m^{2}_{a}}{c^{4}d^{2}_{a}}
+\mathcal{O}(\bar{\gamma}^{2})\nonumber\\
& = & 18.3 \bigg(\frac{\bar{\gamma}}{2.1\times 10^{-5}}\bigg)\bigg(\frac{m_a}{M_{\odot}}\bigg)\bigg(\frac{R_{\odot}}{d_a}\bigg)\quad \mu\mathrm{as}\nonumber\\
  & & +81.1 \bigg(\frac{\bar{\gamma}}{2.1\times 10^{-5}}\bigg) \bigg(\frac{m_a}{M_{\odot}}\bigg)^2\bigg(\frac{R_{\odot}}{d_a}\bigg)^2\quad \mathrm{pas}\nonumber\\
  & & +\mathcal{O}(\bar{\gamma}^{2}).
\end{eqnarray}
Here, $\delta\theta_{\mathrm{GR}}$ is the light deflection equivalently caused by GR and its 1PN and 2PN effects are $\sim 1.75$ as and $\sim 3.50$ $\mu$as respectively, both of which are much greater than the thresholds of future $\sim 1$ nas space missions. $\delta\theta_{\mathrm{STT}}$ represents the discrepancy between STT and GR in the deflection angle.
For the effect of the deviation from GR in STT in the deflection angle which would be measured by experiments conducted in the solar system, the leading effect is 18.3 $\mu$as at the 1PN level. As Damour and Nordtvedt mentioned in Ref. \cite{dam93}, the deviation of $\gamma$ from unity at the levels of $\sim 10^{-7}$ to $\sim 10^{-5}$ might be detected by some new space
projects (e.g. LATOR). The 2PN term in $\delta\theta_{\mathrm{STT}}$ might also be important because it could reach $\sim 0.1$ nas, close to the thresholds of future space-borne experiments.

\subsection{Deflection angle by a 2-point-masses system}

\label{case2pms}

From a 1-point-mass system to an $N$-point-masses system, the spacetime will no longer be spherical symmetric and static. This complicates the problem significantly. We suppose that there are three characteristic length scales: the length scale of the $N$-point-masses system $\mathfrak{L}$, the distance between the light source and the system $\mathfrak{R}_{\lambda}$ and the distance from the observer and the system $\mathfrak{R}_{\mathrm{obs}}$; and two characteristic time scales: the time scale of orbital motions of the $N$-point-masses system $\mathfrak{T}$ and the time scale of the light crossing the system $\mathfrak{T}_{\mathrm{cross}}$.  In general, $\mathfrak{L}$, $\mathfrak{R}_{\lambda}$ and  $\mathfrak{R}_{\mathrm{obs}}$ could be arbitrary, while $\mathfrak{T}$ is determined by the total mass of the system $\mathfrak{M}$ and $\mathfrak{L}$, as well as $\mathfrak{T}_{\mathrm{cross}}\leq\mathfrak{L}/c$.

In what follows, we will consider a specific case that the gravitational system contains only two point masses with body $a$ and $b$ ($N=2$) and  $\mathfrak{L} \gg \mathrm{max}(\mathfrak{R}_{\lambda}, \mathfrak{R}_{\mathrm{obs}})$, and then calculate the resulting deflection angle.

\subsubsection{Light equations}

For this 2-point-masses system with the configuration $\mathfrak{L} \gg \mathrm{max}(\mathfrak{R}_{\lambda}, \mathfrak{R}_{\mathrm{obs}})$, it indicates $\mathfrak{T}_{\mathrm{cross}} \ll \mathfrak{T}$. However, this does not mean that the impact of the orbital motion can be neglected, if this effect firstly emerges at less than or equal to 2PN level. Nevertheless, the calculation including this motion demands a huge amount of work and, thus, we would leave it for our next steps. In this paper, we firstly consider a case of this binary system that the effects of velocity-dependent terms are smaller than those of 2PN static effects, i.e. $F^i_v=0$ in Eq. (\ref{le2}), so that the light equations are
\begin{eqnarray}
\label{2bd}
\ddot{x}^{i}=F^{i}_{1PN}+F^{i}_{2PN}|_{1BE}+F^{i}_{2PN}|_{2BE},
\end{eqnarray}
where the contribution at 1PN  is
\begin{eqnarray}
F^{i}_{1PN}&=&-(1+\gamma)\frac{Gm_{a}}{r^{2}_{a}}\bigg[n^{i}_{a}-2\frac{(n_{a}\dot{x})\dot{x}^{i}}{c^{2}}\bigg]+a\leftrightarrow b,
\end{eqnarray}
the contribution of one-body effect at 2PN is
\begin{equation}
 F^{i}_{2PN}|_{1BE} = \varepsilon^{2}(1+\gamma)\frac{G^{2}m^{2}_{a}}{r^{3}_{a}}\bigg\{2(1+\gamma)n^{i}_{a}
+\frac{(n_{a}\dot{x})^{2}}{c^{2}}n^{i}_{a}
-\frac{(n_{a}\dot{x})\dot{x}^{i}}{c^{2}}\bigg\}+a\leftrightarrow b,
\end{equation}
and the contribution of two-body effect at 2PN is
\begin{eqnarray}
 & & F^{i}_{2PN}|_{2BE}\nonumber\\
 &=&\varepsilon^{2}(1+\gamma)\frac{G^{2}m_{a}m_{b}}{r_{a}}\bigg\{\frac{1}{r_{a}r_{ab}}n^{i}_{a}
-\frac{1}{2r^{2}_{ab}}(n_{a}n_{ab})n^{i}_{a}
+(1-\gamma)\frac{1}{r_{a}r_{b}}n^{i}_{a}\nonumber\\
&&+(1+3\gamma)\frac{1}{r^{2}_{b}}n^{i}_{b}
-\frac{3}{2r^{2}_{ab}}n^{i}_{ab}
-\frac{2}{r_{a}r_{ab}}\frac{(n_{a}\dot{x})\dot{x}^{i}}{c^{2}}
+\frac{(n_{a}n_{ab})}{r^{2}_{ab}}\frac{(n_{a}\dot{x})\dot{x}^{i}}{c^{2}}
-\frac{1}{r^{2}_{ab}}\frac{(n_{ab}\dot{x})\dot{x}^{i}}{c^{2}}\nonumber\\
&&+2(1-\gamma)\frac{1}{r^{2}_{b}}\frac{(n_{b}\dot{x})\dot{x}^{i}}{c^{2}}
-2(1-\gamma)\frac{1}{r_{a}r_{b}}\frac{(n_{a}\dot{x})\dot{x}^{i}}{c^{2}}\bigg\}\nonumber\\
&&+\varepsilon^{2}(1+\gamma)\frac{G^{2}m_{a}m_{b}}{S^{2}_{ab}}\bigg\{-2\frac{1}{r_{ab}}\frac{(n_{ab}\dot{x})(n_{a}\dot{x})+(n_{ab}\dot{x})(n_{b}\dot{x})}{c^{2}}n^{i}_{ab}\nonumber\\
&&-4\frac{1}{S_{ab}}\frac{[(n_{ab}\dot{x})+(n_{b}\dot{x})][(n_{a}\dot{x})+(n_{b}\dot{x})]}{c^{2}}(n^{i}_{ab}-n^{i}_{a})
+\frac{1}{r_{a}}\frac{(n_{ab}\dot{x})+(n_{b}\dot{x})}{c}\bigg[\frac{(n_{a}\dot{x})}{c}n^{i}_{a}-\frac{\dot{x}^{i}}{c}\bigg]\nonumber\\
&&-2\frac{1}{r_{b}}\bigg[\frac{(n_{b}\dot{x})^{2}}{c^{2}}-1\bigg](n^{i}_{ab}-n^{i}_{a})
+2\frac{1}{r_{ab}}\bigg[\frac{(n_{a}\dot{x})\dot{x}^{i}}{c^{2}}+\frac{(n_{b}\dot{x})\dot{x}^{i}}{c^{2}}\bigg]
+\frac{1}{r_{ab}}\frac{(n_{ab}\dot{x})^{2}}{c^{2}}[n^{i}_{a}+n^{i}_{b}]\nonumber\\
&&+2\frac{1}{S_{ab}}\frac{[(n_{ab}\dot{x})-(n_{a}\dot{x})][(n_{ab}\dot{x})+(n_{b}\dot{x})]}{c^{2}}(n^{i}_{a}+n^{i}_{b})
+\frac{1}{r_{b}}\frac{(n_{ab}\dot{x})-(n_{a}\dot{x})}{c}\bigg[\frac{(n_{b}\dot{x})}{c}n^{i}_{b}-\frac{\dot{x}^{i}}{c}\bigg]\nonumber\\
&&-\frac{1}{r_{ab}}(n^{i}_{a}+n^{i}_{b})\bigg\}+a\leftrightarrow b,
\end{eqnarray}
where the symbol $a\leftrightarrow b$ means the same terms but with the labels $a$ and $b$ exchanged.

\subsubsection{Light-ray trajectory}

Following the procedure mentioned in Appendix \ref{lightrayonebody}, we could work out the time derivative of light-ray trajectory $\frac{1}{c}\delta\dot{\bm{x}}(t)$ and compute the deflection angle caused by these two point masses. For $F^{i}_{1PN}$ and $F^{i}_{2PN}|_{1BE}$, we could obtained corresponding trajectory of the light-ray by adopting the iterative method used in Ref. \cite{bru91}. However, in order to integrate $F^{i}_{2PN}|_{2BE}$, we assume the light source, light-ray trajectory and the observer are all  closer to the body $a$ than to the body $b$, i.e. $r_{a}<r_{ab}$ along the light path from end to end. Thus, we could obtain
\begin{eqnarray}
\label{lightresulttwo}
\frac{1}{c}\dot{\bm{x}}(t)&=&\bm{k}+\frac{1}{c}\delta\dot{\bm{x}}_{1PN}(\bm{x}_{N})
+\frac{1}{c}\delta\dot{\bm{x}}_{2PN}(\bm{x}_{N})|_{1BE}+\frac{1}{c}\delta\dot{\bm{x}}_{2PN}(\bm{x}_{N})|_{2BE},
\end{eqnarray}
where
\begin{eqnarray}
\frac{1}{c}\delta\dot{\bm{x}}_{1PN}(\bm{x})&=&-(1+\gamma)\frac{Gm_{a}}{c^{2}r_{a}}
\bigg\{\frac{\bm{k}\times(\bf{r}_{a}\times\bm{k})}{r_{a}-\bm{k}\cdot\bm{r}_{a}}+\bm{k}\bigg\}
-(1+\gamma)\frac{Gm_{b}}{c^{2}r_{b}}
\bigg\{\frac{\bm{k}\times(\bf{r}_{b}\times\bm{k})}{r_{b}-\bm{k}\cdot\bm{r}_{b}}+\bm{k}\bigg\},
\end{eqnarray}

\begin{eqnarray}
 & & \frac{1}{c}\delta\dot{\bm{x}}_{2PN}(\bm{x})|_{1BE}\nonumber\\
 &=&-\frac{1}{4}(1+\gamma)\frac{G^{2}m^{2}_{a}}{c^{4}r^{4}_{a}}(\bm{k}\cdot\bm{r}_{a})\bm{r}_{a}
-\frac{1}{4}(1+\gamma)\frac{G^{2}m^{2}_{b}}{c^{4}r^{4}_{b}}(\bm{k}\cdot\bm{r}_{b})\bm{r}_{b}\nonumber\\
&&+\frac{G^{2}m^{2}_{a}}{c^{4}}\bm{d}_{a}\bigg\{(1+\gamma)^{2}\frac{1}{r_{a}(r_{a}-\bm{k}\cdot\bm{r}_{a})}
\bigg(\frac{2}{r_{a}}+\frac{1}{r_{a}-\bm{k}\cdot\bm{r}_{a}}\bigg)\nonumber\\
&&-\frac{1}{8}(1+\gamma)(7+8\gamma)\frac{1}{d^{2}_{a}}\bigg[\frac{\bm{k}\cdot\bm{r}_{a}}{r^{2}_{a}}
+\frac{1}{d_{a}}\bigg(\frac{\pi}{2}+\arctan\frac{\bm{k}\cdot\bm{r}_{a}}{d_{a}}\bigg)\bigg]\bigg\}\nonumber\\
&&+\frac{G^{2}m^{2}_{a}}{c^{4}r_{a}}\bm{k}\bigg[\frac{1}{4}(1+\gamma)(5+4\gamma)\frac{1}{r_{a}}
-(1+\gamma)^{2}\frac{1}{r_{a}-\bm{k}\cdot\bm{r}_{a}}\bigg]\nonumber\\
&&+\frac{G^{2}m^{2}_{b}}{c^{4}}\bm{d}_{b}\bigg\{(1+\gamma)^{2}\frac{1}{r_{b}(r_{b}-\bm{k}\cdot\bm{r}_{b})}
\bigg(\frac{2}{r_{b}}+\frac{1}{r_{b}-\bm{k}\cdot\bm{r}_{b}}\bigg)\nonumber\\
&&-\frac{1}{8}(1+\gamma)(7+8\gamma)\frac{1}{d^{2}_{b}}\bigg[\frac{\bm{k}\cdot\bm{r}_{b}}{r^{2}_{b}}
+\frac{1}{d_{b}}\bigg(\frac{\pi}{2}+\arctan\frac{\bm{k}\cdot\bm{r}_{b}}{d_{b}}\bigg)\bigg]\bigg\}\nonumber\\
&&+\frac{G^{2}m^{2}_{b}}{c^{4}r_{b}}\bm{k}\bigg[\frac{1}{4}(1+\gamma)(5+4\gamma)\frac{1}{r_{b}}
-(1+\gamma)^{2}\frac{1}{r_{b}-\bm{k}\cdot\bm{r}_{b}}\bigg],
\end{eqnarray}

\begin{eqnarray}
  & & \frac{1}{c}\delta\dot{\bm{x}}_{2PN}(\bm{x})|_{2BE}\nonumber\\
  &=&(1+\gamma)\frac{G^{2}m_{a}m_{b}}{c^{4}r_{a}r_{ab}}\bm{k}\bigg\{-(1+\gamma)\frac{\bm{r}_{ab}\cdot\bm{d}_{a}}{(r_{a}-\bm{k}\cdot\bm{r}_{a})(r_{ab}-\bm{k}\cdot\bm{r}_{ab})}
+(2+\gamma)\bigg\}\nonumber\\
&&+(1+\gamma)\frac{G^{2}m_{a}m_{b}}{c^{4}r^{3}_{ab}}\bm{k}\bigg\{-(1+\gamma)\frac{\bm{r}_{ab}\cdot\bm{d}_{a}}{r_{a}-\bm{k}\cdot\bm{r}_{a}}
-\frac{\bm{r}_{a}\cdot\bm{r}_{ab}}{2r_{a}}-\frac{(\bm{k}\cdot\bm{r}_{a})(\bm{k}\cdot\bm{r}_{ab})}{r_{ab}}\nonumber\\
&&+3\frac{(\bm{k}\cdot\bm{r}_{ab})^{2}}{r^{2}_{ab}}r_{a}
-\frac{1}{2}(7+6\gamma)(\bm{k}\cdot\bm{r}_{ab})\ln\frac{r_{a}+\bm{k}\cdot\bm{r}_{a}}{r_{0a}+\bm{k}\cdot\bm{r}_{0a}}\bigg\}\nonumber\\
&&+(1+\gamma)\frac{G^{2}m_{a}m_{b}}{c^{4}r_{a}r_{ab}}\bm{d}_{a}\bigg\{(1+\gamma)\frac{1}{(r_{a}-\bm{k}\cdot\bm{r}_{a})^{2}(r_{ab}-\bm{k}\cdot\bm{r}_{ab})}\bigg[(\bm{r}_{ab}\cdot\bm{d}_{a})
-(\bm{k}\cdot\bm{r}_{a})(\bm{k}\cdot\bm{r}_{ab})\nonumber\\
&&+(\bm{k}\cdot\bm{r}_{ab})r_{ab}+\frac{1}{2}(\bm{k}\cdot\bm{r}_{ab})r_{a}\bigg]+(3+2\gamma)\frac{1}{r_{a}-\bm{k}\cdot\bm{r}_{a}}
-\frac{1}{2}(1+\gamma)\frac{1}{(r_{a}-\bm{k}\cdot\bm{r}_{a})^{2}}r_{a}\nonumber\\
&&-(1+\gamma)\frac{1}{(r_{a}-\bm{k}\cdot\bm{r}_{a})(r_{ab}-\bm{k}\cdot\bm{r}_{ab})}r_{a}\bigg\}
+(1+\gamma)\frac{G^{2}m_{a}m_{b}}{c^{4}r^{3}_{ab}}\bm{d}_{a}\bigg\{-\frac{1}{2}(3+4\gamma)\frac{\bm{k}\cdot\bm{r}_{ab}}{r_{a}-\bm{k}\cdot\bm{r}_{a}}\nonumber\\
&&+\frac{1}{3}(1+\gamma)\frac{(\bm{k}\cdot\bm{r}_{a})^{2}}{(r_{a}-\bm{k}\cdot\bm{r}_{a})(r_{ab}-\bm{k}\cdot\bm{r}_{ab})^{2}}r_{ab}
+\frac{1}{3}(1+\gamma)\frac{(\bm{k}\cdot\bm{r}_{a})^{2}}{(r_{a}-\bm{k}\cdot\bm{r}_{a})(r_{ab}-\bm{k}\cdot\bm{r}_{ab})}\nonumber\\
&&-\frac{1}{2}\frac{\bm{r}_{a}\cdot\bm{r}_{ab}}{r_{a}(r_{a}-\bm{k}\cdot\bm{r}_{a})}+\bigg[3\frac{(\bm{k}\cdot\bm{r}_{ab})^{2}}{r^{2}_{ab}}-1\bigg]
\ln\frac{r_{a}+\bm{k}\cdot\bm{r}_{a}}{r_{0a}+\bm{k}\cdot\bm{r}_{0a}}\bigg\}\nonumber\\
&&+(1+\gamma)^{2}\frac{G^{2}m_{a}m_{b}}{c^{4}r_{a}r_{ab}}\frac{1}{r_{ab}-\bm{k}\cdot\bm{r}_{ab}}\bm{d}_{b}\bigg[2-\frac{r_{a}}{r_{a}-\bm{k}\cdot\bm{r}_{a}}
-\frac{r_{a}}{r_{ab}-\bm{k}\cdot\bm{r}_{ab}}\ln\frac{r_{a}+\bm{k}\cdot\bm{r}_{a}}{r_{0a}+\bm{k}\cdot\bm{r}_{0a}}\bigg]\nonumber\\
&&+(1+\gamma)^{2}\frac{G^{2}m_{a}m_{b}}{c^{4}r^{2}_{ab}}\frac{1}{(r_{ab}-\bm{k}\cdot\bm{r}_{ab})^{2}}\bm{d}_{b}\bigg[
\frac{\bm{r}_{a}\cdot\bm{r}_{ab}}{r_{a}-\bm{k}\cdot\bm{r}_{a}}
-\frac{3}{2}\frac{(\bm{k}\cdot\bm{r}_{a})(\bm{k}\cdot\bm{r}_{ab})}{r_{a}-\bm{k}\cdot\bm{r}_{a}}+\frac{1}{3}r_{a}\nonumber\\
&&+\frac{1}{2}(\bm{k}\cdot\bm{r}_{ab})\ln\frac{r_{a}+\bm{k}\cdot\bm{r}_{a}}{r_{0a}+\bm{k}\cdot\bm{r}_{0a}}\bigg]
+(1+\gamma)^{2}\frac{G^{2}m_{a}m_{b}}{c^{4}r^{3}_{ab}}\frac{1}{r_{ab}-\bm{k}\cdot\bm{r}_{ab}}\bm{d}_{b}\bigg[
\frac{\bm{r}_{a}\cdot\bm{r}_{ab}}{r_{a}-\bm{k}\cdot\bm{r}_{a}}\nonumber\\
&&-\frac{3}{2}\frac{(\bm{k}\cdot\bm{r}_{a})(\bm{k}\cdot\bm{r}_{ab})}{r_{a}-\bm{k}\cdot\bm{r}_{a}}+\frac{1}{3}r_{a}
+\frac{1}{2}(\bm{k}\cdot\bm{r}_{ab})\ln\frac{r_{a}+\bm{k}\cdot\bm{r}_{a}}{r_{0a}+\bm{k}\cdot\bm{r}_{0a}}\bigg]\nonumber\\
&&+(1+\gamma)\frac{G^{2}m_{a}m_{b}}{c^{4}r^{3}_{ab}}\bm{r}_{ab}\bigg\{4\frac{\bm{k}\cdot\bm{r}_{a}}{r_{ab}}-\frac{\bm{k}\cdot\bm{r}_{a}}{r_{a}}
-3\frac{\bm{k}\cdot\bm{r}_{ab}}{r^{2}_{ab}}r_{a}
-2\frac{(\bm{k}\cdot\bm{r}_{a})(\bm{k}\cdot\bm{r}_{ab})^{2}}{r^{3}_{ab}}\nonumber\\
&&-\frac{d_{a}}{r_{ab}}\bigg(\frac{\pi}{2}+\arctan\frac{\bm{k}\cdot\bm{r}_{a}}{d_{a}}\bigg)
+\frac{1}{2}(1+4\gamma)\ln\frac{r_{a}+\bm{k}\cdot\bm{r}_{a}}{r_{0a}+\bm{k}\cdot\bm{r}_{0a}}\bigg\}\nonumber\\
&&+O\bigg(\frac{1}{c^4}\frac{r_a}{r_{ab}}\bigg),
\end{eqnarray}
where $\bm{r}_{0a}=\bm{r}_{a}(t=t_{0})$ and $r_{0a}=|\bm{r}_{0a}|$. For GR ($\gamma=\beta=1$) and one body case,
Eq. (\ref{lightresulttwo}) will reduce to
the results of Ref. \cite{bru91}.

\subsubsection{Light deflection}

With the definition of a gauge-invariant angle $\theta$ between the directions of two incoming photons (light signals 1 and 2)
and we use that the
position of the photon at the moment $t$ of observation coincides
with the position of the observer so that
\begin{eqnarray}
\bm{x}_{obs}=\bm{x}_{01}+c(t-t_{01})\bm{k}_{1}+\delta\bm{x}_{1}
=\bm{x}_{02}+c(t-t_{02})\bm{k}_{2}+\delta
\bm{x}_{2},
\end{eqnarray}
where $(t_{01},\bm{x}_{01})$ denotes the moment and position of
the light signal $1$ of emission and $(t_{02},\bm{x}_{02})$ for
the light signal $2$ respectively.

By assuming that the initial unit vector of light-ray 1 $\bm{k}_1$ is perpendicular to the line connecting the body $a$ and the $b$, $\bm{r}_{ab}$, and light-ray 2 moves along the line connecting the body $a$ and the observer, which lead to that geometrical relationships that
\begin{eqnarray}
&&\frac{\bm{k}_{1}\cdot\bm{r}_{a}}{r_{a}}\approx\cos\vartheta_{0},\frac{\bm{k}_{2}\cdot\bm{r}_{a}}{r_{a}}=1,\frac{\bm{k}_{1}\cdot\bm{r}_{ab}}{r_{ab}}=0,
\frac{d_{1a}}{r_{a}}\approx\sin\vartheta_{0}\\
&&d_{2a}=0,\frac{\bm{k}_{2}\cdot\bm{d}_{1a}}{d_{1a}}=\sin\vartheta_{0},\frac{\bm{k}_{2}\cdot\bm{r}_{ab}}{r_{ab}}=\sin\vartheta_{0},\bm{k}_{2}=\frac{\bm{r}_{a}}{r_{a}}.
\end{eqnarray}
and letting $d_{1a}\equiv d_{a}$, we could obtain the deflection angle for a static observer as
\begin{eqnarray}
&&\Delta\theta\equiv\theta-\vartheta_{0}\nonumber\\
&=&\bigg(\frac{1+\gamma}{2}\bigg)\frac{4Gm_{a}}{c^{2}d_{a}}\bigg(\frac{1+\cos\vartheta_{0}}{2}\bigg)
+(1+\gamma)\frac{Gm_{a}}{c^{2}r_{ab}}\bigg(1-\frac{\cos\vartheta_{0}}{1-\sin\vartheta_{0}}\bigg)\nonumber\\
&&-2(1+\gamma)^{2}\frac{G^{2}m^{2}_{a}}{c^{4}d^{2}_{a}}\bigg(\frac{1+\cos\vartheta_{0}}{2}\bigg)
+(1+\gamma)^{2}\frac{G^{2}m^{2}_{b}}{c^{4}r^{2}_{ab}}\frac{\sin\vartheta_{0}(1+\sin\vartheta_{0})}{\cos^{2}\vartheta_{0}}\nonumber\\
&&-\frac{1}{8}(1+\gamma)\frac{G^{2}m^{2}_{a}}{c^{4}d^{2}_{a}}\sin\vartheta_{0}\cos\vartheta_{0}
-\frac{1}{8}(1+\gamma)\frac{G^{2}m^{2}_{a}}{c^{4}r^{2}_{ab}}\frac{\sin\vartheta_{0}}{\cos\vartheta_{0}}\nonumber\\
&&+\frac{1}{8}(7+8\gamma)(1+\gamma)\frac{G^{2}m^{2}_{a}}{c^{4}d^{2}_{a}}(\pi-\vartheta_{0})\nonumber\\
&&-\frac{1}{8}(7+8\gamma)(1+\gamma)\frac{G^{2}m^{2}_{a}}{c^{4}r^{2}_{ab}}\bigg[\frac{1}{\cos^{2}\vartheta_{0}}
\bigg(\frac{\pi}{2}+\vartheta_{0}\bigg)-\frac{\pi}{2}\bigg]\nonumber\\
&&-2(1+\gamma)\frac{G^{2}m_{a}m_{b}}{c^{4}r^{2}_{ab}}\sin\vartheta_{0}\cos\vartheta_{0}
-(1+\gamma)^2\frac{G^{2}m_{a}m_{b}}{c^{4}r^{2}_{ab}}\sin\vartheta_{0}\nonumber\\
&&-2(1+\gamma)^{2}\frac{G^{2}m_{a}m_{b}}{c^{4}r_{ab}d_{a}}\bigg(\frac{1+\cos\vartheta_{0}}{2}\bigg)
\bigg(1-\frac{\cos\vartheta_{0}}{1-\sin\vartheta_{0}}\bigg)\frac{\cos\vartheta_{0}}{\sin\vartheta_{0}}\nonumber\\
&&-2(1+\gamma)(2+\gamma)\frac{G^{2}m_{a}m_{b}}{c^{4}r_{ab}d_{a}}\bigg(\frac{1+\cos\vartheta_{0}}{2}\bigg)\nonumber\\
&&+(1+\gamma)^{2}\frac{G^{2}m_{a}m_{b}}{c^{4}r_{ab}d_{a}}\frac{\sin\vartheta_{0}\cos\vartheta_{0}}{1-\sin\vartheta_{0}}\nonumber\\
&&+\mathcal{O}\bigg(\frac{1}{c^4}\frac{r_a}{r_{ab}}\bigg).
\end{eqnarray}
Similar to the case of 1-point-mass system, the 2PN light deflection in a 2-point-masses system with static approximation does \emph{not} depend on PPN parameter $\beta$, neither. One possible reason for this is that $\beta$ usually associates with motion of bodies so that when the system could be treated as a static system due to $\mathfrak{T}_{\mathrm{cross}} \ll \mathfrak{T}$, it would disappear in the deflection angle. However, when $\mathfrak{T}_{\mathrm{cross}} \sim \mathfrak{T}$, we suppose the time evolutions of the gravitational fields, the motion of bodies and PPN parameter $\beta$ would explicitly show up in the expression of deflection angle, and this more complicated issue will be the next move of our investigation.

\subsubsection{Quantitative examples}

If the light-ray just grazes the limb of the body $a$, the angle between light-ray 1 and 2 at the observer $\vartheta_{0}$ could be very close to $0$ and be neglected so that
\begin{eqnarray}
\Delta\theta&=&\bigg(\frac{1+\gamma}{2}\bigg)\frac{4Gm_{a}}{c^{2}d_{a}}\nonumber\\
&&-(1+\gamma)\bigg[2(1+\gamma)-\frac{1}{8}(7+8\gamma)\pi\bigg]\frac{G^{2}m^{2}_{a}}{c^{4}d^{2}_{a}}\nonumber\\
&&-2(1+\gamma)(2+\gamma)\frac{G^{2}m_{a}m_{b}}{c^{4}r_{ab}d_{a}}.
\end{eqnarray}
and it could be written in a more practical form
\begin{equation}
  \label{}
  \Delta\theta = \delta\theta_{\mathrm{GR}} + \delta\theta_{\mathrm{STT}},
\end{equation}
where
\begin{eqnarray}
  \label{thetaGR2be}
  \delta\theta_{\mathrm{GR}} & = & \frac{4Gm_{a}}{c^{2}d_{a}}
-(8-\frac{15}{4}\pi)\frac{G^{2}m^{2}_{a}}{c^{4}d^{2}_{a}}
-12\frac{G^2m_{a}m_{b}}{c^{4}r_{ab}d_{a}}\nonumber\\
  & = & 1.75 \bigg(\frac{m_a}{M_{\odot}}\bigg)\bigg(\frac{R_{\odot}}{d_a}\bigg)\quad \mathrm{arcsecond}\nonumber\\
  & &+3.50\bigg(\frac{m_a}{M_{\odot}}\bigg)^2\bigg(\frac{R_{\odot}}{d_a}\bigg)^2\quad \mu\mathrm{as}\nonumber\\
  & & -51.5 \bigg(\frac{m_b}{m_a}\bigg)\bigg(\frac{m_a}{M_{\odot}}\bigg)^2\bigg(\frac{R_{\odot}}{d_a}\bigg)\bigg(\frac{1\mathrm{AU}}{r_{ab}}\bigg)\quad\mathrm{nas}
\end{eqnarray}
\begin{eqnarray}
  \label{deltatheta2be}
  \delta\theta_{\mathrm{STT}} & = &\bar{\gamma}\frac{2Gm_{a}}{c^{2}d_{a}}
-\bar{\gamma}(8-\frac{31}{8}\pi)\frac{G^{2}m^{2}_{a}}{c^{4}d^{2}_{a}}
-10\bar{\gamma}\frac{G^2m_{a}m_{b}}{c^{4}r_{ab}d_{a}} +\mathcal{O}(\bar{\gamma}^{2})\nonumber\\
& = & 18.3 \bigg(\frac{\bar{\gamma}}{2.1\times 10^{-5}}\bigg)\bigg(\frac{m_a}{M_{\odot}}\bigg)\bigg(\frac{R_{\odot}}{d_a}\bigg)\quad \mu\mathrm{as}\nonumber\\
  & & +81.1 \bigg(\frac{\bar{\gamma}}{2.1\times 10^{-5}}\bigg) \bigg(\frac{m_a}{M_{\odot}}\bigg)^2\bigg(\frac{R_{\odot}}{d_a}\bigg)^2\quad \mathrm{pas}\nonumber\\
  & & - 0.902 \bigg(\frac{\bar{\gamma}}{2.1\times 10^{-5}}\bigg)\bigg(\frac{m_b}{m_a}\bigg)\bigg(\frac{m_a}{M_{\odot}}\bigg)^2\bigg(\frac{R_{\odot}}{d_a}\bigg)\bigg(\frac{1~\mathrm{AU}}{r_{ab}}\bigg)\quad\mathrm{pas}\nonumber\\
  & & +\mathcal{O}(\bar{\gamma}^{2}).
\end{eqnarray}
If we ignore the mass of body $b$, the results Eqs. (\ref{thetaGR2be}) and (\ref{deltatheta2be}) could return to the 1-point-mass case Eqs. (\ref{thetaGR1be}) and (\ref{deltatheta1be}).

For experiments conducted in the Solar System, these results could be applied to some cases that only the Sun and the Jupiter need to be considered. If we suppose a light-ray passing the limb of the Sun, letting the body $a$ be the Sun in Eqs. (\ref{thetaGR2be}) and (\ref{deltatheta2be}), then the 2PN deflection angle of $\delta\theta_{\mathrm{GR}}$ contains $3.50$ $\mu$as caused by the Sun itself and $-9.46$ pico-arcsecond (pas) due to the coupling term of the Sun and the Jupiter. For the effect of the deviation from GR at the 2PN level in $\delta\theta_{\mathrm{STT}}$, it consists of $81.1$ pas from the Sun itself and $-0.166$ femto-arcsecond (fas) caused by the coupling effect of the Sun and the Jupiter, both of which are below the thresholds of future space missions.

In another case that a light-ray just grazes the limb of the Jupiter, letting the body $a$ be the Jupiter in Eqs. (\ref{thetaGR2be}) and (\ref{deltatheta2be}), the deflection angle at 2PN level of $\delta\theta_{GR}$ consists of $0.302$ nas by the Jupiter and $-0.0921$ nas by the coupling of the Sun and the Jupiter. The 2PN contributions in $\delta\theta_{\mathrm{STT}}$ are made of $7.00$ fas by the Jupiter and $-1.61$ fas by the coupling of the Sun and the Jupiter.

One of the major differences of this work from previous works is abandoning the hierarchized hypothesis, which leads to the 2-body effects in the light deflection at 2PN level. Although these effects in the 2PN light deflection is at least 1 order of magnitude lower than the thresholds of future $\sim$nas experiments in the Solar System, their contributions in $\delta\theta_{\mathrm{GR}}$ and $\delta\theta_{\mathrm{STT}}$ could reach $-0.515$ $\mu$as and $-9.02$ pas respectively for an imaginary observer in a binary system with two $\sim 1$ $M_{\odot}$ Sun-like stars and separation by $\sim 0.1$ AU, both of which increase 3 orders of magnitude more than the system with mass ratio $\sim10^{-3}$ (such as $M_{\mathrm{Jupiter}}/M_{\odot}$) and the same separation. Hence, if this observer would be able to measure $\bar{\gamma}$ down to $\sim 10^{-6}$, the 2-body effect in $\delta\theta_{\mathrm{GR}}$ had to be considered, because the third term in $\delta\theta_{\mathrm{GR}}$ and the first term in $\delta\theta_{\mathrm{STT}}$ are comparable with the level of $\sim 1$ $\mu$as.

An even more extreme case would be replacing one Sun-like star in above hypothetical example with a neutron star with radius $\sim$10 km, which could make the 2PN 2-body effects in $\delta\theta_{\mathrm{GR}}$ and $\delta\theta_{\mathrm{STT}}$ raise to $-35.9$ mas and $-0.627$ $\mu$as.

\section[]{Conclusions and discussion}

In this paper, we mainly focus on the scalar-tensor theory and obtain its 2PN metric. To investigate its next to the leading order contributions in the light propagation and related measurements in the near zone of a gravitational system, we reduce the metric to its 2PN form for light by supposing an $N$-point-masses system. In doing so, at 2PN level, we abandon the hierarchized hypothesis and do not assume two PPN parameters $\gamma$ and $\beta$ to be unity, which makes our work differ from previous works. It is found that although there exist $\gamma$ and $\beta$ in the 2PN
metric, only $\gamma$ appears in the 2PN equations of light.

As one simple example of applications for future experiments, a gauge-invariant angle
between the directions of two incoming photons for a differential
measurement is investigated after the light trajectory is obtained in a static and spherically symmetric spacetime. It shows the deviation from GR $\delta\theta_{\mathrm{STT}}$ does \emph{not} depend on $\beta$ even at 2PN level in this circumstance, which is consistent  with previous results.

A more complicated application is light deflection in a 2-point-masses system. We consider a case that the light crossing time is much less than the time scale of its orbital motion, $\mathfrak{T}_{\mathrm{cross}} \ll \mathfrak{T}$, and thus treat it as a ``static'' system. The 2-body effect at 2PN level originating from relaxing the hierarchized hypothesis is calculated. Our analysis shows the 2PN 2-body effect in the Solar System is one order of magnitude less than future $\sim 1$ nas experiments, while this effect could be comparable with 1PN component of $\delta\theta_{\mathrm{STT}}$ in a binary system with two Sun-like stars and separation by $\sim 0.1$ AU if an experiment would be able to measure $\bar{\gamma}$ down to $\sim 10^{-6}$.

Our next move is to relax the static approximation for the 2-point-masses system with $\mathfrak{T}_{\mathrm{cross}} \sim \mathfrak{T}$ and investigate the time evolution pattern of the 2PN light deflection. We suppose the PPN parameter $\beta$ would appear in the result.

\begin{acknowledgments}
The authors are grateful to an anonymous referee whose comments are very helpful to the
final version of this article.
We are thankful to the International Space Science Institute for hospitality and accommodation.
The authors especially would like to thank Professor Tian-Yi Huang of Nanjing University for his fruitful discussions.
The work of X.-M. Deng is
funded by the Natural Science Foundation of China under
Grant No. 11103085. The work of Y. Xie is supported by the China Scholarship Council Grant No.
2008102243, the National Natural Science Foundation of China Grant Nos.
10973009 and 11103010, the Fundamental Research Program of Jiangsu
Province of China No. BK2011553 and the Research Fund for the Doctoral
Program of Higher Education of China No.20110091120003. This project/publication was made possible through the support
of a grant from the John Templeton Foundation. The opinions expressed in
this publication are those of the authors and do not necessarily reflect
the views of the John Templeton Foundation. The funds from John Templeton
Foundation were awarded in a grant to The University of Chicago which also
managed the program in conjunction with National Astronomical Observatories,
Chinese Academy of Sciences. X.-M. Deng appreciates the support from the group of Almanac and Astronomical
Reference Systems in the Purple Mountain Observatory of China.
\end{acknowledgments}

\appendix

\section{Evolution equations of metric coefficients of second order post-Newtonian approximation}

In this appendix, we will give the metric coefficients of the scalar-tensor theory at the 2PN order as follows.

\subsection{$N$ and $\overset{(2)}{\zeta}$}

The equations for $N$ and $\overset{(2)}{\zeta}$ are
\begin{equation}
\label{N}  \Box N=-8\pi G \sigma,
\end{equation}
\begin{equation}
\Box\overset{(2)}{\zeta}=-4(1-\gamma)\pi G \sigma,
\end{equation}
where $\Box$ is the D'Alembert operator in the Minkowski spacetime
and
\begin{eqnarray}
    \label{gamma}
    \gamma & \equiv & \frac{\omega_0+1}{\omega_0+2},\\
    \label{G}
    G & \equiv & \frac{2}{\phi_{0}(1+\gamma)}.
\end{eqnarray}
And we can see $G=1/\phi_{0}$ when $\gamma=1$.

\subsection{$H_{ij}$}

The equation for $H_{ij}$ yields
\begin{equation}
    \label{eqHij}
    \Box H_{ij} = -8\gamma\pi G \sigma\delta_{ij},
\end{equation}
which gives $H_{ij}\equiv V\delta_{ij}$. Then, we obtain
\begin{equation}
    \label{eqV}
    \Box V=-8\gamma\pi G \sigma.
\end{equation}
Furthermore, we have $V=\gamma N$ by comparison with Eqs. (\ref{N})
and (\ref{eqV}).

\subsection{$L_{i}$}

The equation for $L_{i}$ is
\begin{equation}
    \label{Li}
    \Box L_i = 8(\gamma+1) \pi G \sigma_i.
\end{equation}

\subsection{$L$ and $\overset{(4)}{\zeta}$}

The equations for $L$ and $\overset{(4)}{\zeta}$ are
\begin{eqnarray}
    \label{L}
    \Box L & = & -8\pi G \sigma\bigg(V-2N+\frac{4\beta-\gamma-3}{\gamma-1}\overset{(2)}{\zeta}\bigg)-8 (\gamma-1)\pi G\sigma_{kk}\nonumber\\
    &&-\frac{1}{2}N_{,k}N_{,k}-\frac{1}{2}N_{,k}V_{,k}-\overset{(2)}{\zeta}_{,k}N_{,k}-\frac{4(\beta-1)}{(\gamma-1)^2}\overset{(2)}{\zeta}_{,k}\overset{(2)}{\zeta}_{,k},
\end{eqnarray}
\begin{eqnarray}
    \label{}
    \Box \overset{(4)}{\zeta} & = & -4\pi G \sigma \bigg[(\gamma-1)(N-V)+\frac{8(\beta-1)}{\gamma-1}\overset{(2)}{\zeta}\bigg]-8(\gamma-1)\pi G \sigma_{kk}\nonumber\\
    &&+\frac{1}{2}\overset{(2)}{\zeta}_{,k}N_{,k}-\frac{1}{2}\overset{(2)}{\zeta}_{,k}V_{,k}+\frac{4(1-\beta)}{(\gamma-1)^2}\overset{(2)}{\zeta}_{,k}\overset{(2)}{\zeta}_{,k},
\end{eqnarray}
where
\begin{eqnarray}
    \label{beta}
    \beta & \equiv & 1+\frac{\omega_1}{(2\omega_0+3)(2\omega_0+4)^2}.
\end{eqnarray}

\subsection{$Q_{ij}$}

The equation for $Q_{ij}$ reads as
\begin{eqnarray}
    \label{}
    \Box Q_{ij} & = & -8(1+\gamma) \pi G \sigma_{ij}\nonumber\\
    &&-\frac{1}{2}N_{,i}N_{,j}+VN_{,ij}-VV_{,ij}-\frac{1}{2}V_{,i}V_{,j}\nonumber\\
    &&-\overset{(2)}{\zeta}N_{,ij}-\frac{1}{2}\overset{(2)}{\zeta}_{,i}N_{,j}-\frac{1}{2}\overset{(2)}{\zeta_{,j}}N_{,i}+\frac{1}{2}\overset{(2)}{\zeta}_{,i}V_{,j}+\frac{1}{2}\overset{(2)}{\zeta}_{,j}V_{,i}\nonumber\\
    &&-2V\overset{(2)}{\zeta}_{,ij}+\overset{(2)}{\zeta}V_{,ij}+2\overset{(2)}{\zeta}\overset{(2)}{\zeta}_{,ij}+\frac{2(2\gamma-1)}{\gamma-1}\overset{(2)}{\zeta}_{,i}\overset{(2)}{\zeta}_{,j}\nonumber\\
    &&+\delta_{ij}\bigg\{ +8\pi G \sigma \bigg[+\gamma(N-2V)+\frac{4\beta-4+\gamma^2-\gamma}{\gamma-1}\overset{(2)}{\zeta}\bigg]+16\gamma\pi G \sigma_{kk} \nonumber\\
    &&\phantom{+\delta_{ij}\bigg\{}+\frac{1}{2}N_{,k}V_{,k}+\frac{1}{2}V_{,k}V_{,k}-\overset{(2)}{\zeta}_{,k}V_{,k}+\frac{4(\beta-1)}{(\gamma-1)^2}\overset{(2)}{\zeta}_{,k}\overset{(2)}{\zeta}_{,k}\bigg\}.
\end{eqnarray}

\subsection{$Q_{i}$}

The equation for $Q_{i}$ is
\begin{eqnarray}
    \label{Qi}
    \Box Q_i & = &+8(\gamma+1)\pi G \sigma_i \bigg(2V-N-\overset{(2)}{\zeta}\bigg)+8\pi G \sigma L_i\nonumber\\
    &&+\frac{1}{2}N_{,i}N_{,t}+N_{,i}V_{,t}+\frac{1}{2}V_{,i}N_{,t}+\frac{1}{2}VN_{,it}+V_{,i}V_{,t}-\frac{1}{2}VV_{,it}\nonumber\\
    &&-\frac{1}{2}N_{,k}L_{i,k}+\frac{1}{2}N_{,ik}L_k-N_{,k}L_{k,i}+\frac{1}{2}V_{,k}L_{i,k}-\frac{1}{2}V_{,ik}L_k-V_{,k}L_{k,i}\nonumber\\
    &&-\frac{1}{2}\overset{(2)}{\zeta}_{,t}N_{,i}-\frac{1}{2}\overset{(2)}{\zeta}N_{,it}-\overset{(2)}{\zeta}_{,it}V+\frac{3}{2}\overset{(2)}{\zeta}_{,t}V_{,i}+\frac{1}{2}\overset{(2)}{\zeta}V_{,it}\nonumber\\
    &&-L_{i,k}\overset{(2)}{\zeta}_{,k}-\overset{(2)}{\zeta}_{,ik}L_k+\frac{(3\gamma-1)}{\gamma-1}\overset{(2)}{\zeta}_{,i}\overset{(2)}{\zeta}_{,t}
    +\overset{(2)}{\zeta}\overset{(2)}{\zeta}_{,it}.
\end{eqnarray}

\subsection{$Q$ and $\overset{(6)}{\zeta}$}

The equations for $Q$ and $\overset{(6)}{\zeta}$ are
\begin{eqnarray}
    \label{Q}
    \Box Q & = & -8\pi G \sigma \bigg\{+N^2-2NV-2L-\frac{4\beta-\gamma-3}{\gamma-1}\bigg(2N-V\bigg)\overset{(2)}{\zeta}\nonumber\\
    &&\phantom{-8\pi G \sigma \bigg\{}+\bigg[1-\frac{\iota}{(\gamma-1)^2}-\frac{4(\beta-1)(\gamma^2+8\beta-2\gamma-7)}{(\gamma-1)^3}\bigg]\overset{(2)}{\zeta}^2\nonumber\\
    &&\phantom{-8\pi G \sigma \bigg\{}+\frac{4\beta-\gamma-3}{\gamma-1}\overset{(4)}{\zeta}\bigg\}\nonumber\\
    &&-8\pi G \sigma_{kk}\bigg[(2\gamma-1)V+(2-\gamma)N-\frac{\gamma^2+8\beta-2\gamma-7}{\gamma-1}\overset{(2)}{\zeta}\bigg]\nonumber\\
    &&+16\pi G \sigma_kL_k-\frac{1}{2}NN_{,k}N_{,k}+\frac{1}{2}VN_{,k}V_{,k}\nonumber\\
    &&+\frac{3}{2}N_{,t}N_{,t}+\frac{3}{2}V_{,t}V_{,t}+N_{,t}V_{,t}+NN_{,tt}+VN_{,tt}\nonumber\\
    &&-N_{,k}L_{k,t}-L_kN_{,kt}-V_{,kt}L_k-N_{,k}L_{,k}-\frac{1}{2}V_{,k}L_{,k}\nonumber\\
    &&+N_{,k}Q_{kl,l}+N_{,kl}Q_{kl}-\frac{1}{2}N_{,l}Q_{kk,l}-L_{k,l}L_{l,k}+L_{l,k}L_{l,k}\nonumber\\
    &&+2\overset{(2)}{\zeta}_{,t}V_{,t}-2\overset{(2)}{\zeta}_{,kt}L_k-\overset{(2)}{\zeta}_{,k}L_{,k}\nonumber\\
    &&+\frac{2(\gamma^2-\gamma+2\beta-2)}{(\gamma-1)^2}\overset{(2)}{\zeta}_{,t}\overset{(2)}{\zeta}_{,t}+\frac{4(\beta-1)}{(\gamma-1)^2}N\overset{(2)}{\zeta}_{,k}\overset{(2)}{\zeta}_{,k}+\overset{(2)}{\zeta}\overset{(2)}{\zeta}_{,k}N_{,k}\nonumber\\
    &&+\frac{2\iota}{(\gamma-1)^3}\overset{(2)}{\zeta}\overset{(2)}{\zeta}_{,k}\overset{(2)}{\zeta}_{,k}+\frac{4(\beta-1)(\gamma^2+8\beta-2\gamma-7)}{(\gamma-1)^4}\overset{(2)}{\zeta}\overset{(2)}{\zeta}_{,k}\overset{(2)}{\zeta}_{,k}\nonumber\\
    &&-\overset{(4)}{\zeta}_{,k}N_{,k}-\frac{8(\beta-1)}{(\gamma-1)^2}\overset{(2)}{\zeta}_{,k}\overset{(4)}{\zeta}_{,k},
\end{eqnarray}
and
\begin{eqnarray}
    \label{}
    \Box\overset{(6)}{\zeta}& = & -4\pi G \sigma \bigg[+(\gamma-1)(NV+L)-\frac{8(\beta-1)}{\gamma-1}\overset{(2)}{\zeta}(N-V)\nonumber\\
    &&\phantom{-4\pi G \sigma \bigg[}-\frac{2(32\beta^2+\iota\gamma-64\beta-\iota+32)}{(\gamma-1)^3}\overset{(2)}{\zeta}^2+\frac{8(\beta-1)}{\gamma-1}\overset{(4)}{\zeta}\bigg]\nonumber\\
    &&-4\pi G \sigma_{kk}\bigg[(\gamma-1)(3V-N)-\frac{16(\beta-1)}{\gamma-1}\overset{(2)}{\zeta}\bigg]-8\pi G (\gamma-1)\sigma_kL_k\nonumber\\
    &&+\frac{1}{2}N\overset{(2)}{\zeta}_{,k}N_{,k}+\frac{1}{2}V\overset{(2)}{\zeta}_{,k}V_{,k}-2\overset{(2)}{\zeta}_{,kt}L_k-\overset{(2)}{\zeta}_{,k}L_{k,t}+\frac{1}{2}\overset{(2)}{\zeta}_{,k}L_{,k}\nonumber\\
    &&+\overset{(2)}{\zeta}_{,k}Q_{kl,l}+\overset{(2)}{\zeta}_{,lk}Q_{lk}-\frac{1}{2}\overset{(2)}{\zeta}_{,l}Q_{kk,l}+(N+V)\overset{(2)}{\zeta}_{,tt}\nonumber\\
    &&+\frac{4\beta+2\gamma-\gamma^2-5}{(\gamma-1)^2}\overset{(2)}{\zeta}_{,t}\overset{(2)}{\zeta}_{,t}+\frac{32(\beta-1)^2+2\iota(\gamma-1)}{(\gamma-1)^4}\overset{(2)}{\zeta}\overset{(2)}{\zeta}_{,k}\overset{(2)}{\zeta}_{,k}\nonumber\\
    &&-\frac{8(\beta-1)}{(\gamma-1)^2}\overset{(2)}{\zeta}_{,k}\overset{(4)}{\zeta}_{,k}+\frac{1}{2}\overset{(4)}{\zeta}_{,k}N_{,k}-\frac{1}{2}\overset{(4)}{\zeta}_{,k}V_{,k},
\end{eqnarray}
where
\begin{eqnarray}
    \label{iota}
    \iota & \equiv & \frac{1}{2}\frac{(\gamma-1)^4}{\gamma+1}\omega_2.
\end{eqnarray}
In general relativity, $\iota=0$ \cite{b33}.

\section{Application in a static, spherically symmetric spacetime}

\label{lightrayonebody}

\subsection{Light-ray trajectory}

In this appendix, we will consider the gravitational field outside a static,
spherically symmetric body and mainly focus on the parameters in 2PN level in several gravity theories.
The trajectory of the light-ray can be obtained through
integrating the following equation
\begin{eqnarray}
\ddot{\bm{x}}&=&-(1+\gamma)\frac{Gm_{a}}{r^{2}_{a}}\bigg[\bm{n}_{a}-2\frac{(n_{a}\dot{x})\dot{\bm{x}}}{c^{2}}\bigg]\nonumber\\
&&+\varepsilon^{2}(1+\gamma)\frac{G^{2}m^{2}_{a}}{r^{3}_{a}}\bigg\{\bigg[2(1+\gamma)+\frac{(n_{a}\dot{x})^{2}}{c^{2}}\bigg]\bm{n}_{a}
-\frac{(n_{a}\dot{x})\dot{\bm{x}}}{c^{2}}\bigg\},
\end{eqnarray}
by adopting the iterative method used in Ref. \cite{bru91}. We assume the unperturbed light-ray as follows
\begin{eqnarray}
\bm{x}_{N}&=&\bm{x}_{0}+c(t-t_{0})\bm{k},
\end{eqnarray}
where $k^{i}$ is a unit vector representing the light
direction at $t=-\infty$, $t_{0}$ is an instant on the light path
and $x^{i}_{0}$ is the position of photon at $t_{0}$. The
instant $t_{0}$ can be arbitrarily chosen, for example the emission
or the observation instant. The photon's
coordinates can be written as sum of perturbations on
$x^{i}_{N}$:
\begin{eqnarray}
\bm{x}(t)&=&\bm{x}_{N}+\delta \bm{x}\equiv \bm{x}_{N}+\delta
\bm{x}_{1PN}+\delta\bm{x}_{2PN}.
\end{eqnarray}
And we use the following assumption for
motion of the bodies
\begin{eqnarray}
\bm{r}_{A}(t)&=&\bm{x}(t)-\bm{y}_{a}=\bm{x}(t)-\bm{y}_{a}(t_{a}),
\end{eqnarray}
where $t_{a}$ is the moment of the closest approach between the body
$a$ and the unperturbed light ray. $\bm{y}_{a}(t_{a})$ means the position of $a$-th body ($\bm{y}_{a}$) is evaluated at the time $t_{a}$. After these, we obtain the results with the method as \cite{bru91}
\begin{eqnarray}
\label{lightresult1}
\frac{1}{c}\dot{\bm{x}}(t)&=&\bm{k}+\frac{1}{c}\delta\dot{\bm{x}}_{1PN}(\bm{x}_{N})
+\frac{1}{c}\delta\dot{\bm{x}}_{2PN}(\bm{x}_{N}),
\end{eqnarray}
\begin{eqnarray}
\label{liahtresult2}
\bm{x}(t)&=&\bm{x}_{N}(t)+\bigg[\delta\bm{x}_{1PN}(\bm{x}_{N})-\delta\bm{x}_{1PN}(\bm{x}_{0})\bigg]
+\bigg[\delta\bm{x}_{2PN}(\bm{x}_{N})-\delta\bm{x}_{2PN}(\bm{x}_{0})\bigg].
\end{eqnarray}
It is worthy of note that the 2PN terms in our solution actually have two sources, direct
and indirect. The direct part comes from the 2PN order itself. The indirect part comes
from the 1PN terms when the 1PN solution is iterated into itself in order to attain a 2PN
accuracy, namely, we substitute $\bm{x}_{N}+\delta\bm{x}_{1PN}$ into the trajectory of the light-ray in the 1PN
approximation. And
\begin{eqnarray}
\frac{1}{c}\delta\dot{\bm{x}}_{1PN}(\bm{x})&=&-(1+\gamma)\frac{Gm_{a}}{c^{2}r_{a}}
\bigg\{\frac{\bm{k}\times(\bf{r}_{a}\times\bm{k})}{r_{a}-\bm{k}\cdot\bm{r}_{a}}+\bm{k}\bigg\},\\
\frac{1}{c}\delta\dot{\bm{x}}_{2PN}(\bm{x})&=&-\frac{1}{4}(1+\gamma)\frac{G^{2}m^{2}_{a}}{c^{4}r^{4}_{a}}(\bm{k}\cdot\bm{r}_{a})\bm{r}_{a}
+\frac{G^{2}m^{2}_{a}}{c^{4}}\bm{d}_{a}\bigg\{(1+\gamma)^{2}\frac{1}{r_{a}(r_{a}-\bm{k}\cdot\bm{r}_{a})}
\bigg(\frac{2}{r_{a}}+\frac{1}{r_{a}-\bm{k}\cdot\bm{r}_{a}}\bigg)\nonumber\\
&&-\frac{1}{8}(1+\gamma)(7+8\gamma)\frac{1}{d^{2}_{a}}\bigg[\frac{\bm{k}\cdot\bm{r}_{a}}{r^{2}_{a}}
+\frac{1}{d_{a}}\bigg(\frac{\pi}{2}+\arctan\frac{\bm{k}\cdot\bm{r}_{a}}{d_{a}}\bigg)\bigg]\bigg\}\nonumber\\
&&+\frac{G^{2}m^{2}_{a}}{c^{4}r_{a}}\bm{k}\bigg[\frac{1}{4}(1+\gamma)(5+4\gamma)\frac{1}{r_{a}}
-(1+\gamma)^{2}\frac{1}{r_{a}-\bm{k}\cdot\bm{r}_{a}}\bigg],
\end{eqnarray}
and
\begin{eqnarray}
\delta\bm{x}_{1PN}(\bm{x})&=&-(1+\gamma)\frac{Gm_{a}}{c^{2}}\bigg\{\frac{\bm{k}\times(\bm{r}_{a}\times\bm{k})}{r_{a}-\bm{k}\cdot\bm{r}_{a}}
+\bm{k}\ln(r_{a}+\bm{k}\cdot\bm{r}_{a})\bigg\},\\
\delta\bm{x}_{2PN}(\bm{x})&=&\frac{1}{8}(1+\gamma)\frac{G^{2}m^{2}_{a}}{c^{4}r^{2}_{a}}\bm{r}_{a}+\frac{G^{2}m^{2}_{a}}{c^{4}}\bm{k}\bigg[
-\frac{1}{8}(1+\gamma)(7+8\gamma)\frac{1}{d_{a}}\arctan\frac{\bm{k}\cdot\bm{r}_{a}}{d_{a}}\nonumber\\
&&-(1+\gamma)^{2}\frac{1}{r_{a}-\bm{k}\cdot\bm{r}_{a}}\bigg]
+\frac{G^{2}m^{2}_{a}}{c^{4}}\bm{d}_{a}\bigg\{(1+\gamma)^{2}\frac{1}{(r_{a}-\bm{k}\cdot\bm{r}_{a})^{2}}\nonumber\\
&&-\frac{1}{8}(1+\gamma)(7+8\gamma)\frac{\bm{k}\cdot\bm{r}_{a}}{d^{3}_{a}}\bigg[\frac{\pi}{2}+\arctan\frac{\bm{k}\cdot\bm{r}_{a}}{d_{a}}\bigg]\bigg\},
\end{eqnarray}
where $\bm{d}_{a}=\bm{k}\times(\bm{r}_{a}\times\bm{k})$
is an impact parameter to represent the closest distance between the
unperturbed light ray and body $a$, $d_{a}=|\bm{d}_{a}|$. For GR ($\gamma=\beta=1$),
Eqs. (\ref{lightresult1}) and (\ref{liahtresult2}) will reduce to
the results of Ref. \cite{bru91}.

\subsection{Light deflection}

In some practical astronomical measurements, a differential
measurement is more powerful. This concept is employed by
LATOR mission \citep{tur04,tur04b} through a skinny triangle formed
by two spacecrafts and the International Space Station. Thus, we construct a gauge-invariant
angle $\theta$ between the directions of two incoming photons based
on \cite{bru91,b26}. It reads
\begin{equation}
\label{cosdef}
\cos\theta=\frac{h_{\alpha\beta}K^{\alpha}_{1}K^{\beta}_{2}}
{\sqrt{h_{\alpha\beta}K^{\alpha}_{1}K^{\beta}_{1}}\sqrt{h_{\alpha\beta}K^{\alpha}_{2}K^{\beta}_{2}}},
\end{equation}
where the spatial projection operator is
\begin{equation}
h_{\alpha\beta}=g_{\alpha\beta}+u_{\alpha}u_{\beta},
\end{equation}
which projects the two incoming photons onto the hypersurface
orthogonal to the observer's four-velocity $u^{\alpha}\equiv
\mathrm{d}x^{\alpha}/c\mathrm{d}\tau$ and $K^{\alpha}_{1}\equiv
\mathrm{d}x^{\alpha}_{1}(t)/\mathrm{d}t$, and $K^{\beta}_{2}\equiv
\mathrm{d}x^{\beta}_{2}(t)/\mathrm{d}t$ are the tangent vectors of
the paths $x^{\alpha}_{1}(t)$ and $x^{\beta}_{2}(t)$ of the two
incoming photons. Then, for a static observer, we obtain
\begin{eqnarray}
\label{theta}
\theta&=&\vartheta_{0}+(1+\gamma)\frac{Gm_{a}}{c^{2}r_{a}\sin\vartheta_{0}}\bigg\{
\frac{\bm{k}_{1}\cdot[\bm{k}_{2}\times(\bm{r}_{a}\times\bm{k}_{2})]}{r_{a}
-\bm{k}_{2}\cdot\bm{r}_{a}}+
\frac{\bm{k}_{2}\cdot[\bm{k}_{1}\times(\bm{r}_{a}\times\bm{k}_{1})]}{r_{a}
-\bm{k}_{1}\cdot\bm{r}_{a}}\bigg\}\nonumber\\
&&-(1+\gamma)^{2}\frac{G^{2}m^{2}_{a}}{c^{4}\sin\vartheta_{0}}\bigg[
\frac{\bm{k}_{2}\cdot\bm{d}_{1a}}{d^{3}_{1a}}\bigg(1+\frac{\bm{k}_{1}\cdot\bm{r}_{a}}{r_{a}}\bigg)+
\frac{\bm{k}_{1}\cdot\bm{d}_{2a}}{d^{3}_{2a}}\bigg(1+\frac{\bm{k}_{2}\cdot\bm{r}_{a}}{r_{a}}\bigg)\bigg]\nonumber\\
&&-\frac{1}{8}(1+\gamma)\frac{G^{2}m^{2}_{a}}{c^{4}r^{2}_{a}}\bigg[\frac{\bm{k}_{1}\cdot\bm{r}_{a}}{d_{1a}}+
\frac{\bm{k}_{2}\cdot\bm{r}_{a}}{d_{2a}}\bigg]\nonumber\\
&&+\frac{1}{8}(7+8\gamma)(1+\gamma)\frac{G^{2}m^{2}_{a}}{c^{4}\sin\vartheta_{0}}\bigg\{
\frac{\bm{k}_{1}\cdot\bm{d}_{2a}}{d^{3}_{2a}}\bigg[\frac{\pi}{2}+\arctan\bigg(\frac{\bm{k}_{2}\cdot\bm{r}_{a}}{d_{2a}}\bigg)\bigg]\nonumber\\
&&+\frac{\bm{k}_{2}\cdot\bm{d}_{1a}}{d^{3}_{1a}}\bigg[\frac{\pi}{2}+\arctan\bigg(\frac{\bm{k}_{1}\cdot\bm{r}_{a}}{d_{1a}}\bigg)\bigg]
\bigg\},
\end{eqnarray}
where
\begin{eqnarray}
\vartheta_{0}=\arccos(\bm{k}_{1}\cdot\bm{k}_{2}),
\end{eqnarray}
and we use that the
position of the photon at the moment $t$ of observation coincides
with the position of the observer so that
\begin{eqnarray}
\label{constrant}
\bm{x}_{obs}=\bm{x}_{01}+c(t-t_{01})\bm{k}_{1}+\delta\bm{x}_{1}
=\bm{x}_{02}+c(t-t_{02})\bm{k}_{2}+\delta
\bm{x}_{2},
\end{eqnarray}
where $(t_{01},\bm{x}_{01})$ denotes the moment and position of
the light signal $1$ of emission and $(t_{02},\bm{x}_{02})$ for
the light signal $2$ respectively. In Eq. (\ref{theta}), $\bm{d}_{1a}=\bm{k}_{1}\times(\bm{r}_{a}\times\bm{k}_{1})$ and
is an impact parameter to represent the closest distance between the
unperturbed light ray 1 and $a$-th body. Similarly, $\bm{d}_{2a}=\bm{k}_{2}\times(\bm{r}_{a}\times\bm{k}_{2})$ and
is an impact parameter to represent the closest distance between the
unperturbed light ray 2 and $a$-th body.

Furthermore, we assume that one of the two light rays moves along the line
connecting the body $a$ and the observer, namely source 2. This means the
impact parameter for source 2 is zero so that
$d_{2a}=|\bm{k}_{2}\times(\bm{r}_{a}\times\bm{k}_{2})|=0$,
$\bm{k}_{2}\cdot\bm{d}_{1a}/d_{1a}=\sin\vartheta_{0}$,
$\bm{k}_{1}\cdot\bm{r}_{a}/r_{a}\simeq\cos\vartheta_{0}$, $d^{2}_{1a}=r^{2}_{a}-(\bm{k}_{1}\cdot\bm{r}_{a})^{2}$,
$\bm{k}_{1}\cdot\bm{r}_{a}/d_{1a}\approx\pi/2$ and $d_{1a}/r_{a}=0$. Then, we obtain Eq. (\ref{def}).

\newpage %Just because of unusual number of tables stacked at end
\bibliography{apssamp}% Produces the bibliography via BibTeX.

\end{document}